\pdfoutput=1
\documentclass[a4paper,12pt]{article}
\usepackage{ifpdf}
\ifpdf
\usepackage[pdftex]{graphicx}
\usepackage[usenames,dvipsnames]{color}
\else
\usepackage{graphicx}
\usepackage[usenames,dvips]{color}
\fi
\usepackage{amsmath}
\usepackage{amssymb}
\usepackage[colorlinks=true]{hyperref}   
\usepackage{caption}
\usepackage{paper}

\numberwithin{equation}{section}

\def\SchwOnePLogTimeMax{100.0}

\def\SchwOnePLogArtViscosity{0.08}

\def\SchwOnePLogNumNode{2048}

\def\SchwGeodNumNode{2048}
\def\SchwGeodMass{1.0}
\def\SchwGeodLength{800.0}
\def\SchwGeodLxx{0.01}
\def\SchwGeodRx{0.25}

\def\LapseSlope{-0.54215}
\def\LapseSlopeExact{-0.54433}
\def\LapseSlopeError{0.4}

\begin{document}

\title{An Einstein-Bianchi system for\\[5pt]
       Smooth Lattice General Relativity. I. \\[5pt]
       The Schwarzschild spacetime.}
\author{%
Leo Brewin\\[10pt]%
School of Mathematical Sciences\\%
Monash University, 3800\\%
Australia}
\date{10-Nov-2010}
\date{16-Jan-2011}
\date{26-Jan-2011}
\reference{Preprint: arXiv:1101.3171}

\maketitle

\begin{abstract}
\noindent
The second Bianchi identity can be recast as an evolution equation for the Riemann curvatures.
Here we will report on such a system for a vacuum static spherically symmetric spacetime. This
is the first of two papers. In the following paper we will extend the ideas developed here to
general vacuum spacetimes. In this paper we will demonstrate our ideas on a Schwarzschild
spacetime and give detailed numerical results. For suitable choices of lapse function we find
that the system gives excellent results with long term stability.
\end{abstract}

\section{Introduction}
\label{sec:intro}

Despite a slow start, hyperbolic formulations of the Einstein equations have in recent times
become the system of choice for numerical relativity.

The confidence afforded to hyperbolic systems is borne out not just by the recent success in
numerical relativity \cite{pretorius:2005-01,campanelli:2006-01,baker:2006-01} but also from
their strict mathematical underpinnings (which guarantees that future evolutions exist and
that they depend smoothly on the initial data, these are key aspects of the theorems that
demonstrate the stability of the system, see \cite{reula:1998-01} for details).

One of the earlier hyperbolic formulations was given by Friedrich \cite{friedrich:1996-01} in
which he used the second Bianchi identities to evolve the Weyl curvatures in-situ with the
metric. This idea has been extended by many other authors
\cite{anderson:1999-01,estabrook:1997-01,van-elst:1997-01,jantzen:2009-01} and the resulting
equations are commonly referred to as an Einstein-Bianchi system.

Yet despite their mathematical elegance and the virtues that this would bestow upon a
numerical code there seems to be have been very few numerical applications employing an
Einstein-Bianchi system (though see \cite{van-putten:1996-01,van-putten:1997-01}).

In this paper we will report on a simple Einstein-Bianchi system adapted to a discrete
lattice for static spherically symmetric spacetimes. We were lead to this formulation not by
way of Friedrich's paper but rather as a direct extension of our own ideas developed in an
earlier series of papers \cite{brewin:2009-05,brewin:2009-04,brewin:2002-01,brewin:1998-02}.
In those papers we used the spatial form of the second Bianchi identities (\ie the second
Bianchi identity for the 3-metric) to compute the 3-Riemann curvatures across a Cauchy
surface. This device proved to be the key element in obtaining accurate and stable evolutions
of the initial data.

Our longer term intention is to employ an Einstein-Bianchi system to evolve a 3-dimensional
lattice. This will require not only evolution equations for the legs of the lattice, such as
those given in \cite{brewin:2009-04}, but also evolution equations for the curvatures. This
will be the subject of the second paper in this series.

For the simple case presented here we find that the system works very well. The evolutions are
stable, though this depends on the choice of the lapse function, see sections
(\ref{sec:ArtVisco}) and (\ref{sec:Slicing}). We also find that the constraints are well
behaved (they appear to grow linearly with time and converge to zero as the lattice is
refined, see section (\ref{sec:CodeTests})).

As this paper borrows heavily from two of our previous papers, which we refer to as \PaperI\
and \PaperII, we will skip over many of the derivations and arguments assuming instead that
the reader is familiar with the material in \PaperI\ and \PaperII.

\section{Spherically symmetric spacetimes}
\label{sec:SphericalSpacetime}

In this paper we will be constructing lattice approximations to the Schwarzschild spacetime in
various slicings. In each case the continuum metric can be written in the form
\begin{equation}
ds^2 = -N(r,t)^2 dt^2 + A(r,t)^2 dr^2 + B(r,t)^2 d\Omega^2
\label{eqn:SphericalMetric}
\end{equation}
for some set of functions $N(r,t)$, $A(r,t)$, $B(r,t)$ and where $d\Omega^2=d\theta^2 +
\sin^2\theta d\phi^2$ is the metric of the unit 2-sphere. We have introduced this coordinate
form of the continuum metric simply as a precursor to the introduction of the lattice. As we
shall soon see, we will use the coordinate lines and their local tangent vectors as a scaffold
on which to build the lattice, after which we will have no further need for the coordinates
(indeed we could dispense with the coordinates altogether at the possible expense of the
clarity of the exposition).

Consider a local orthonormal tetrad built from the future pointing unit normal $t^\mu$ to a
typical Cauchy surface and three unit vectors $m^\mu_x$, $m^\mu_y$ and $m^\mu_z$ where
$m^\mu_z$ is parallel to the radial axis (see figure
(\ref{fig:LatticeLocal})). These basis vectors are also tangent
vectors to the coordinate axes of a local Riemann normal frame. We will use this tetrad to
record the frame components of the extrinsic and Riemann curvatures on the lattice. Our
notation, which we borrow from \PaperI, will be to use script characters to denote frame
quantities, thus $\Kxx:=K_{\mu\nu} m^\mu_x m^\nu_x$ while 
$\Rtztz:=R_{\mu\alpha\nu\beta}  t^\mu m^\alpha_z t^\nu m^\beta_z$.
Also, to avoid an overflow of symbols, we will allow $\Lxx$ and $\Lzz$ to represent both the
length of the corresponding leg as well the leg itself.

In this class of spacetimes, and on this tetrad, we know that the extrinsic curvature is
diagonal and that a basis for the non-trivial Riemann curvatures is given by
\begin{gather*}
\Rx,\quad \Rz,\quad \Rtxtx,\quad \Rtztz,\quad \Rtxxz
\end{gather*}
Now using $R_{\mu\nu}=R^{\alpha}{}_{\mu\alpha\nu}$ and $R=g^{\mu\nu}R_{\mu\nu}$ we find
\begin{align}
\Rtz &= -2\Rtxxz\\[2pt]
\Rtt &= \Rtztz + 2\Rtxtx\\[2pt]
\Rzz &= -\Rtztz + 2\Rz\\[2pt]
\Rxx &= -\Rtxtx + \Rx + \Rz\\[2pt]
\R &= -4\Rtxtx - 2\Rtztz + 2\left(\Rx+2\Rz\right)
\end{align}
while the non-trivial vacuum Einstein equations yield
\begin{equation}
\Rtz = \Rtt = \Rzz = \Rxx = 0
\end{equation}
Combining the above shows that we can express all of the non-trivial Riemann curvatures solely
in terms of $\Rx$ and $\Rz$, namely
\begin{align}
\Rtxxz & = 0\label{eqn:Rtxxz}\\[2pt]
\Rtztz & = 2\Rz\label{eqn:Rtztz}\\[2pt]
\Rtxtx & = \Rx + \Rz\label{eqn:Rtxtx}
\end{align}
In obtaining these relations we used
$g^{\mu\nu} = -t^\mu t^\nu + m^\mu_x m^\nu_x + m^\mu_y m^\nu_y + m^\mu_z m^\nu_z$.

Note that $\Rx$ and $\Rz$ are not independent for the simple equation $R=0$
leads to
\begin{equation}
0 = \Rx + 2\Rz
\end{equation}
We will use this equation as a check on our numerical integrations (see section
(\ref{sec:CodeTests}) for more details).

\section{The lattice}
\label{sec:LatticeDescribe}

The symmetries in the Schwarzschild spacetime allows us to use a very simple ladder-like
structure for the lattice, as indicated in figure (\ref{fig:SchwarzLattice}). One way to
imagine the construction of the lattice is to consider the coordinate mesh generated by
setting $t={}$constant and $\theta=\pi/2$ in the coordinate form of the metric in
(\ref{eqn:SphericalMetric}). Then the rungs of the ladder are generated by small increments in
$\phi$ leading to $\Lxx\approx B\Delta\phi$ while the side rails would coincide with two
radial curves (\ie $\phi={}$constant) with $\Lzz\approx A\Delta r$. Clearly, specifying all of
the $\Lxx$ and $\Lzz$ is equivalent to specifying the metric components $A(r,t)$ and $B(r,t)$.
Note that throughout this paper we treat the $\Lxx$ and $\Lzz$ as continuous functions of
time.

We will label the nodes from $0$ to $\Nnode$ and on the few occasions where we need to
discuss more than one leg at a time we will write $\iLxx_i$ to denote an $\Lxx$ leg at node
$i$. In the same way $\iLzz_i$ will denote the $\Lzz$ that joins the nodes $i$ and $i+1$.
Similar notation will be used for other data on the lattice.

The initial data (as described in section (\ref{sec:InitialData})) are constructed in a way
that guarantees reflection symmetry at the throat (which is always tied to node 0).

In our computer code we extend our lattice a small way over the throat, by including the nodes
-3 to -1, so that we can readily impose the reflection symmetries (by simply copying data
across the throat, at no point do we independently evolve any of the data to the left of the
throat).

\section{The evolution equations}
\label{sec:EvolveEqtns}

Our present task is to develop evolution equations for the leg-lengths, the extrinsic
curvatures and, the principle innovation in this paper, evolution equations for the Riemann
curvatures.

A simple derivation of the evolution equations for our lattice can be obtained from a general
pair of equations developed in \PaperII. There it was shown that the first and second
variations of arc lengths can be written in a form remarkably similar to the ADM equations,
namely
\begin{align}
\DLsqDt &= -2 N K_{\mu\nu} \Delta x^\mu_{ij} \Delta x^\nu_{ij} + \BigO{L^3}
\label{eqn:ADMDLijb}\\[10pt]
\DDLsqDt &= 2 N_{|\alpha\beta} \Dxij^\alpha \Dxij^\beta
\label{eqn:ADMDLijc}\\
              & \quad+ 2 N\left(  K_{\mu\alpha}K^\mu{}_\beta 
                                - R_{\mu\alpha\nu\beta} t^\mu t^\nu \right)
                      \Dxij^\alpha \Dxij^\beta + \BigO{L^3}\notag
\end{align}
Note that in the following we will ignore the leading error terms $\BigO{L^3}$. Applying these
equations to the two legs $\Lxx$ and $\Lzz$ of our spherically symmetric lattice leads
immediately to
\begin{align}
\DLxx & = -N \Kxx \Lxx\label{eqn:DLxx}\\[5pt]
\DLzz & = -N \Kzz \Lzz\label{eqn:DLzz}\\[5pt]
\DKxx & = -\dNdxx + N\left( \Rtxtx + \Kxx^2\right)\label{eqn:DKxx}\\[5pt]
\DKzz & = -\dNdzz + N\left( \Rtztz + \Kzz^2\right)\label{eqn:DKzz}
\end{align}
The last part of the picture is to provide evolution equations for the Riemann curvatures,
$\Rx$ and $\Rz$. The basic idea is to rearrange the terms in the Bianchi identities to isolate
the time derivatives while estimating the spatial derivatives from data imported from
neighbouring cells. The calculations are straight-forward but a bit tedious to present here so
we defer the full details to the Appendix. This leads to the following evolution equations
\begin{align}
\DRxDt &= 2N\Kxx\left( 2\Rx  + \Rz\right)\label{eqn:EvolveRx}\\[5pt]
\DRzDt &= 3N\Kxx\Rz + N\Kzz\left(\Rx + 2\Rz\right)\label{eqn:EvolveRz}
\end{align}

The Riemann curvatures $\Rx$ and $\Rz$ would normally not be \emph{evolved} but rather
\emph{derived} from the lattice data such as the leg lengths $\Lxx$ and $\Lzz$. In \PaperI\ we
used (discrete versions of) the geodesic deviation equation and the spatial Bianchi
identity\footnote{Here $z$ is the proper distance measured up the middle of the lattice \ie
along a trajectory that passes through the mid-points of each $\Lxx$}
\begin{align}
0 &= \DDLxxDz + {}^{3}\Rz \Lxx\label{eqn:GeodDev}\\[5pt]
0 &= \frac{d\left(\LLxx {}^{3}\Rx\right)}{dz} - {}^{3}\Rz\DLLxxDz\label{eqn:Bianchi}
\end{align}
to compute the 3-dimensional Riemann curvatures ${}^{3}\Rx$ and ${}^{3}\Rz$ on the lattice. 
In raising the $\Rx$ and $\Rz$ to dynamical variables on the lattice we are forced to view
equations (\ref{eqn:GeodDev},\ref{eqn:Bianchi}) as constraints on the lattice data. In section
(\ref{sec:CodeTests}) we shall present discretised versions of these constraints which we will
later use to check the quality of our numerical results.

The one remaining constraint is the standard momentum constraint (see \PaperI\ for details)
\begin{equation}
0 = \DLKxxDz - \Kzz{\DLxxDz}
\label{eqn:Momentum}
\end{equation}

\subsection{Artificial viscosity}
\label{sec:ArtVisco}

Our numerical experiments (which we will present shortly) showed that the future evolutions
can be subject to high-frequency instabilities. This was seen to occur only in the cases where
the lapse function was controlled by its own evolution equation (\eg as in Harmonic slicing).
For such cases we found that stability could be recovered with the addition of an artificial
viscosity term to the evolution equations.

Let $W$ be any one of the dynamical variables, $\Lxx$, $\Lzz$, $\Kxx$, $\Kzz$. Then the
artificial viscosity is introduced by the addition of a simple dissipation term to the
evolution equation for $W$. After some experimentation we settled on the following form
\begin{equation}
\frac{dW_i}{dt} = \frac{d{\bar W}_i}{dt} 
                + \mu N_i\left( \frac{W_{i+1}-W_i}{\iLzz_i}
                               -\frac{W_i-W_{i-1}}{\iLzz_{i-1}}\right)
\end{equation}
where $d{\bar W}_i/dt$ is the right hand side of the original evolution equation
(\ref{eqn:DLxx}--\ref{eqn:DKzz}) and $\mu$ is a (small) constant. Other choices were tried but
this form seemed to produce stable evolutions for the longest periods of time. Note that we do
not add the dissipation terms to the evolution equations for the curvatures (doing so seemed
to make no difference to the evolutions and had no effect in controlling the instabilities).

How should $\mu$ be chosen? We need to choose it large enough to ensure that the evolution is
stable over a given time interval while also keeping it sufficiently small so as to not effect
the large scale features of the numerical solution. By trial and error we found that setting
$\mu=\SchwOnePLogArtViscosity$ worked well for evolutions to $t=\SchwOnePLogTimeMax$ using
$\Nnode=\SchwOnePLogNumNode$ nodes. We also found that as the number of nodes was increased we
had to make a proportionate increase in $\mu$ to maintain the same quality of the evolution
over the same time interval. That is $\mu =\BigO{\Nnode}$.

The dissipation term is easily seen to be a finite difference approximation to $\mu \Lzz d^2
W/dz^2$ and thus it may appear to be like a Kreiss-Oliger term that vanishes in the
continuum limit. However, since we are forced to set $\mu=\BigO{\Nnode}$ and as
$\Lzz=\BigO{1/\Nnode}$ we see that the term $\mu\Lzz$ is approximately constant, say $\mu'$,
and thus the dissipation term is actually of the form $\mu' d^2W/dz^2$. This is a standard
dissipation term commonly used in hydrodynamic simulations and it does not vanish in the
continuum limit.

\section{Initial data}
\label{sec:InitialData}

The initial data on the lattice are the $\Lxx$, $\Lzz$, $\Kxx$, $\Kzz$, $\Rx$ and $\Rz$ at
each node of the lattice. Their time symmetric initial values were set by a combination of the
Hamiltonian constraint, the geodesic deviation equation and the Bianchi identities. A full
account of the choices made in coming to the equations described below can be found \PaperI.
Here we will just quote the relevant equations simply to provide explicit details of how we
constructed our initial data.

To ensure that the initial data is time symmetric we set $\Kxx=0$ and $\Kzz=0$.

The $\Lzz$ were set according to the method of Bernstein, Hobill and Smarr
\cite{bernstein:1989-01} using $n=\SchwGeodNumNode$ on a grid of length $\SchwGeodLength{}m$.
The ADM mass, $m$, was set to be $\SchwGeodMass$ and the $\Lxx$, $\Rx$ and $\Rz$, for 
$i=1,2,3\dots n$, were set according to
\begin{align}
\iLxx_{i} &= \iLxx_{i-1} + \frac{\iLzz_{i-1}}{\iLzz_{i-2}}
                           \left(\iLxx_{i-1}-\iLxx_{i-2}\right)\notag\\
          &\quad  - \frac{1}{2}\iLzz_{i-1}\left(\iLzz_{i-1}+\iLzz_{i-2}\right)
                    \left(\Lxx\Rz\right)_{i-1}\label{eqn:LxxInit}\\[5pt]
\iRz_i &= \iRz_{i-1}\left(\frac{5\iLLxx_{i-1}
         -\iLLxx_{i}}{5\iLLxx_{i}
         -\iLLxx_{i-1}}\right)\label{eqn:SchwRzInit}\\[5pt]
\iRx_i &= -2\iRz_i\label{eqn:SchwRxInit}
\end{align}
At the reflection symmetric throat (\ie at node 0) we set $\iLxx_0 =\SchwGeodLxx$ and
$\iRx_0=-2\iRz_0=\SchwGeodRx$.

\section{Results}
\label{sec:Results}

In all of our results we used a 4-th order Runge-Kutta integrator with the time step set equal
to $1/2$ the smallest $\Lzz$ on the lattice (which happens to be $\iLzz_0$).

\subsection{Slicing conditions}
\label{sec:Slicing}

We ran our code for eight distinct slicing conditions, some were set by simple algebraic
expressions while others involved differential operators.

We made four choices for the algebraic slicings,
\begin{align}
N&=\exp(-2\Kxx)\label{eqn:LapseAlgebraicA}\\[5pt]
N&=\frac{20\Lxx}{1+20\Lxx}\label{eqn:LapseAlgebraicB}\\[5pt]
N&=\exp(-\Rx)\label{eqn:LapseAlgebraicC}\\[5pt]
N&=\frac{1}{1+\Rx}\label{eqn:LapseAlgebraicD}
\end{align}
and three choice for the differential slicings,
\begin{align}
&\text{$1+\log$ slicing}&&\frac{dN}{dt} = -2N K\label{eqn:LapseOnePLog}\\[5pt]
&\text{Harmonic slicing}&&\frac{dN}{dt} = -N^2 K\label{eqn:LapseHarmonic}\\[5pt]
&\text{Maximal slicing}&&\nabla^2 N = {}^3 RN\label{eqn:LapseMaximal}
\end{align}

The eighth slicing condition was the simple case of geodesic slicing $N=1$.

The algebraic slicings were introduced after our early explorations with the differential
lapses, all of which developed high-frequency instabilities after a short time (well before
$t=100$). The algebraic slicings did not require any artificial viscosity and performed
remarkably well, showing no signs of instabilities to at least $t=1000$ (excluding the lapse
(\ref{eqn:LapseAlgebraicB}) which hits the singularity at $t\approx32$). We have not run our
codes beyond $t=1000$ so we can not comment its stability for $t>1000$.

\subsection{Code tests and results}
\label{sec:CodeTests}

We subjected our code to many of the tests used in \PaperI, such as the time at which geodesic
slicing hits the singularity, the rate at which the lapse at the throat collapses in maximal
slicing and the constancy of $\Lxx$ on the horizon. The results for these various slicings are
shown in figures (\ref{fig:GeodesicLapse}--\ref{fig:Lx3RxErr}). All of the results are as
expected. For the geodesic slicing the code crashes at approximately one time step short of
the singularity. The familiar exponential collapse of the lapse for maximal slicing is evident
in figure (\ref{fig:LapseProfiles}). In this case it is known that the lapse at the throat
should behave as $N\sim\beta\exp(\alpha t)$ for $t\rightarrow\infty$ with
$\alpha=-(2/3)^{(3/2)}\approx\LapseSlopeExact$, see \cite{beig:1998-01}. We estimated the
slope of $\ln N$ vs $t$ from our numerical data to be $\LapseSlope$ which agrees with the
exact value to within $\LapseSlopeError$ percent.

We also have a new test obtained by a simple combination of the evolution equations.
From equations (\ref{eqn:EvolveRx},\ref{eqn:EvolveRz}) we find that
\begin{equation}
\frac{d\left( \Rx  + 2\Rz \right)}{dt}= 2N\left(2\Kxx+\Kzz\right)\left( \Rx  + 2\Rz \right)
\label{eqn:ConserveHamiltonian}
\end{equation}
and as $0=\Rx+2\Rz$ on the initial slice (by construction, see (\ref{eqn:SchwRxInit})) we
conclude that $0=\Rx+2\Rz$ for all time. This is not surprising, our evolution equations for
the curvatures are based on the Bianchi identities and these are guaranteed to preserve the
constraints. If we now set $0=\Rx+2\Rz$ in (\ref{eqn:EvolveRz}) and combine the result with
(\ref{eqn:DLxx}) we find
\begin{equation}
0=\frac{d\Lxx^3 \Rx}{dt}
\label{eqn:ConserveLx3Rx}
\end{equation}
This gives us a new test of our code, that the quantity $\Lxx^3 \Rx$ should be constant
throughout the evolution. Importantly this applies to all slicing conditions. In figure
(\ref{fig:Lx3RxErr}) we have plotted the fractional variations in $\Lxx^3 \Rx$ for two choices
of slicings. We see that the errors for the $1+\log$ slicing are much larger than those for
the algebraic slicing which we attribute to the use of an artificial viscosity. This last
claim is easily checked by varying the artificial viscosity parameter $\mu$. We find that the
errors in $\Lxx^3 \Rx$ varies linearly with $\mu$. Note that in obtaining equation
(\ref{eqn:ConserveLx3Rx}) we have ignored the higher order error terms that would arise if we
had carried through the $\BigO{L^3}$ truncation error from (\ref{eqn:ADMDLijb}). Thus even if
we set $\mu=0$ we can expect some variation of $\Lxx^3 \Rx$ over time (though this variation
should vanish more rapidly than $\BigO{L^3}$).

We also have three constraint equations, namely the geodesic deviation equation
(\ref{eqn:GeodDev}), the 3-dimensional Bianchi identity (\ref{eqn:Bianchi}) and the momentum
constraint (\ref{eqn:Momentum}). The discrete form of these equations are
\begin{align}
P &= \frac{D^2\Lxx}{Dz^2} + {}^{3}\Rz \Lxx\label{eqn:DiscreteGeodDev}\\[5pt]
Q &= \frac{{\tilde D}\left(\LLxx {}^{3}\Rx\right)}{Dz} 
     - {}^{3}{\tilde{\cal R}_{xzxz}}
              \frac{{\tilde D}\Lxx^2}{Dz}\label{eqn:DiscreteBianchi}\\[5pt]
M &= \frac{D\left(\Lxx\Kxx\right)}{Dz} - \Kzz\frac{D\Lxx}{Dz}\label{eqn:DiscreteMomentum}
\end{align}
where ${\tilde{\cal R}_{xzxz}}$ is the average of $\Rz$ across $\Lzz$ while $D/Dz$ and
${\tilde D}/Dz$ are discrete derivative operators defined as follows. For a typical smooth
function $f(z)$ sampled at the grid points $z_i$ we define
\def\p{{\hbox{\small\tt +}}}
\def\o{{\hbox{\small\tt o}}}
\def\m{{\hbox{\small\tt -}}}
\def\fp{f^\p}
\def\fo{f^\o}
\def\fm{f^\m}
\def\Lzzp{\Lzz^\p}
\def\Lzzo{\Lzz^\o}
\def\Lzzm{\Lzz^\m}
\begin{align}
\left(\frac{{\tilde D}f}{Dz}\right)_i &:=\frac{\fp-\fo}{\Lzzo}\\[5pt]
\left(\frac{Df}{Dz}\right)_i &:=
   \frac{1}{\Lzzo+\Lzzm}\left( \Lzzm\left(\frac{\fp-\fo}{\Lzzo}\right)
                              +\Lzzo\left(\frac{\fo-\fm}{\Lzzm}\right) \right)\\[5pt]
\left(\frac{D^2f}{Dz^2}\right)_i &:=
   \frac{2}{\Lzzo+\Lzzm}\left( \frac{\fp-\fo}{\Lzzo}
                              -\frac{\fo-\fm}{\Lzzm} \right)
\end{align}
where we have introduced the superscripts $\p$, $\o$ and $\m$ to denote quantities at the grid
points $z_{i+1}$, $z_i$ and $z_{i-1}$ respectively. Note that the sample points $z_i$ are
constructed from the lattice $\Lzz$ by the recurrence relation $z_{i+1}=z_i + (\Lzz)_i$ with
$z_0=0$. In this notation we have ${\tilde{\cal R}}_{xyxy}:=(\Rx^{\p}+\Rx^{\o})/2$. Finally we
note that the 3-curvatures can be computed from the 4-curvatures by way of the Gauss equation,
\begin{align}
{}^{3}\Rx &= \Rx - \Kxx^2\\
{}^{3}\Rz &= \Rz - \Kxx \Kzz
\end{align}
Ideally we would like to see $P=Q=M=0$ but in reality we expect $P_i$, $Q_i$ and $M_i$ to be
non-zero but small. This is indeed what we observe, see figure (\ref{fig:Lx3RxErr}). We also
computed a crude estimate of the rate of convergence (of $Q$, $P$ and $M$ to zero at a fixed
time) by running our code twice, once with $\Nnode=2048$ and once with $\Nnode=1024$ and then
forming suitable ratios of the constraints at the horizon. In this manner we estimated, in the
absence of artificial viscosity, that $P=\BigO{\Nnode^{-4}}$, $Q=\BigO{\Nnode^{-2}}$ and
$M=\BigO{\Nnode^{-3}}$ while the addition of artificial viscosity degraded the convergence to
$P=\BigO{\Nnode^{-1}}$, $Q=\BigO{\Nnode^{-1}}$ and $M=\BigO{\Nnode^{-2}}$.

We also tried setting ${\tilde D}/Dz:= D/Dz$ and ${\tilde{\cal R}}_{xyxy}:=\Rx^{\o}$ in the
discrete Bianchi constraint but this lead to a reduction in the rate of convergence. The form
of the discrete Bianchi constraint as given above (\ref{eqn:DiscreteBianchi}) is readily seen
\cite{brewin:2002-01} to be a second-order accurate estimate to the continuum Bianchi identity
at the centre of the leg $\Lzz$.

One might ask why we have not included the Hamiltonian constraint in our code tests. The
simple answer is that it is trivially satisfied by our discrete equations. This follows from
the discussion surrounding equation (\ref{eqn:ConserveHamiltonian}) where we showed that
$0=\Rx + 2\Rz$ for all time. It follows that the Hamiltonian $H:=G_{\mu\nu}t^\mu t^\mu$ will
also vanish for all time. Note that this analysis was based on our discrete equations, not on
the continuum equations. We did indeed check that our code maintained $0=\Rx+2\Rz$ 
throughout the evolution.

\clearpage

\appendix

\section{Bianchi identities}
\label{sec:Bianchi}

Here we will use the Bianchi identities to obtain evolution equations for the two curvatures
$\Rxyxy$ and $\Rxzxz$. We will follow the method given in our earlier paper
\cite{brewin:2002-01} in which we used data imported from the neighbouring computational cells
to estimate (by a finite difference approximation) the various derivatives required in the
Bianchi identities. We will employ Riemann normal coordinates\footnote{For more details on
Riemann normal coordinate see \cite{brewin:2009-03} and the references cited therein.}, one
for each computational cell, with the origin centred on the central vertex and the coordinate
axes aligned with those described in section (\ref{sec:SphericalSpacetime}), see also figure
(\ref{fig:LatticeLocal}). In these coordinates, the metric in a typical computational cell is
given by
\begin{equation*}
g_{\mu\nu}(x) = g_{\mu\nu} 
              - \frac{1}{3} \rmanb x^\alpha x^\beta 
              - \frac{1}{6} \drmanbg x^\alpha x^\beta x^\gamma
              + \BigO{L^4}\\[5pt]
\end{equation*}
where $L$ is a typical length scale for the computational cell and $\gmn$ and $\rmanb$ are
constant throughout the computational cell. A convenient choice for $\gmn$ is
$\diag(-1,1,1,1)$ (such a choice can always be made by suitable gauge transformations within
the class of Riemann normal frames). In this case the frame components $\Rxyxy$ and $\Rxzxz$
reduce to the coordinate components $\rxyxy$ and $\rxzxz$ respectively. A further advantage of
using Riemann normal coordinates is that at the origin, where the connection vanishes,
covariant derivatives reduce to partial derivatives.


The two Bianchi identities that we need are
\begin{align}
0 &= \dRxyxyt - \dRtyxyx + \dRtxxyy\label{eqn:RiemEvolA}\\
0 &= \dRxzxzt - \dRtzxzx + \dRtxxzz\label{eqn:RiemEvolD}
\end{align}
This pair of equations contains 4 spatial derivatives each of which we will estimate by a
finite difference approximation. But in order to do so we must first have a sampling of the 4
curvatures at a cluster of points near and around the central vertex. Our simple ladder-like
lattice, with its collection of computational cells along one radial axis, would allow us to
compute only the $z$ partial derivatives. For the $x$ and $y$ derivatives we will need to
extend the lattice along the $x$ and $y$ axes. In short we need a truly 3 dimensional
lattice. Fortunately this is rather easy to do for this spacetime. We can use the spherical
symmetry of the Schwarzschild spacetime to clone copies of the ladder (by spherical rotations)
so that a typical central vertex of the parent ladder-lattice becomes surrounded by 4 copies
of itself. It has two further nearby vertices, fore and aft along the radial axis, that are
themselves central vertices of neighbouring cells in the original ladder-like lattice. 
In figure (\ref{fig:LatticeCloned}) we display an $xz$ slice of the cloned lattice.

We now need the coordinates of all six of the neighbouring vertices. This would require a
solution of 
\begin{equation}
\Lsqij = g_{\mu\nu} \left(x_j^\mu-x_i^\mu\right) \left(x_j^\nu-x_i^\nu\right)
            - \frac{1}{3} \rmanb x^\mu_i x^\nu_i x^\alpha_j x^\beta_j
         + \BigO{L^5}
\label{eqn:RNCLsq}
\end{equation}
for the $x^\mu_i$ for given values for the $\Lij$ and $\rmanb$. However, as we are only going
to use these coordinates to construct transformation matrices which will in turn multiply the
Riemann curvatures, it is sufficient to solve (\ref{eqn:RNCLsq}) using a flat metric. Note
that the above equations can only be used to compute (in fact estimate) the spatial
coordinates of the vertices. For the time coordinates we can appeal to the smoothness of the
underlying metric\footnote{If $(t,x^i)$ are the coordinates for a local Riemann normal frame,
then a smooth Cauchy surface through $(0,0,0,0)$ is described locally by $2t = -K_{ij} x^i
x^j$ and as each $x^i=\BigO{L}$ we also have $t=\BigO{L^2}$.} to argue that for each vertex
$t=\BigO{L^2}$. The result is that the typical central vertex, with coordinates $(0,0,0,0)$,
will have 6 neighbouring central vertices with coordinates as per Table (\ref{tbl:Coords}).

\bgroup
\def\A#1{\hbox to 1.0cm{\hfill$#1$\hfill}}
\def\B#1{\hbox to 1.0cm{\hfill$#1$\hfill}}
\def\C#1{\hbox to 1.0cm{\hfill$#1$\hfill}}
\begin{table}[ht]
\def\H{\vrule height 14pt depth  7pt width 0pt} 
\def\m{\vrule height  0pt depth 10pt width 0pt} 
\def\M{\vrule height 15pt depth 10pt width 0pt}
\begin{center}
\begin{tabular}{cccccccc}
\hline
\H&Vertex&&$t$&$x$&$y$&$z$&\\
\hline
\M&0 && (\A{0},&\A{0},&\B{0},&\C{0})\\
\m&1 && (\A{0},&\A{\Lxx},&\B{0},&\C{0})\\
\m&2 && (\A{0},&\A{0},&\B{\Lyy},&\C{0})\\
\m&3 && (\A{0},&\A{-\Lxx},&\B{0},&\C{0})\\
\m&4 && (\A{0},&\A{0},&\B{-\Lyy},&\C{0})\\
\m&5 && (\A{0},&\A{0},&\B{0},&\C{\Lzz})\\
\m&6 && (\A{0},&\A{0},&\B{0},&\C{-\Lzz})\\
\hline
\end{tabular}
\end{center}
\caption{%
Riemann normal coordinates, to $\BigO{L^2}$, of the central vertex and its 6 immediate
neighbours. These coordinates were computed using a flat space approximation.}
\label{tbl:Coords}
\end{table}
\egroup

This accounts for the structure of our lattice but what values should we assign to the
curvatures at the newly created vertices? Let $(A)_{PQ}$ denote the value of a quantity $A$ at
the vertex $P$ in the local Riemann normal frame for vertex $Q$. Since our spacetime is
spherically symmetric we can assert that
\begin{equation*}
(A)_{00} = (A)_{11} = (A)_{22} = (A)_{33} = (A)_{44}
\end{equation*}
Then the idea that we will import data from neighbouring cells can be expressed as
\begin{equation*}
(A)_{PQ} = (U)_{PQ} (A_{PP})
\end{equation*}
where $(U)_{PQ}$ is the transformation matrix, evaluated at $P$, from the Riemann normal frame
of $P$ to that of $Q$. This matrix will be composed of spatial rotations and boosts.

To get the correct estimates for the first partial derivatives we need only compute $U$ to
terms linear in the leg-lengths.

As an example, let us suppose we wished to compute $v^\mu{}_{,x}$ for a spherically symmetric
vector field $v$ on the lattice. We start with $(v)_{10}=(U)_{10}(v)_{11}$ and
\begin{equation*}
(U)_{10} = (B)_{10}(R)_{10}
\end{equation*}
where $(R)_{10}$ represents a rotation in the $x-y$ plane and $(B)_{10}$ a boost in the $t-x$
plane. Note that as we are working only to linear terms in the lattice scale the order in
which we perform the rotation and boost does not matter. Thus we have
\begin{align*}
(R)_{10} &=
\begin{bmatrix}
1&0&0&0\\
0&\phantom{-}\cos\alpha&\sin\alpha&0\\
0&-\sin\alpha&\cos\alpha&0\\
0&0&0&1
\end{bmatrix}\\[5pt]
(B)_{10} &=
\begin{bmatrix}
\cosh\beta&\sinh\beta&0&0\\
\sinh\beta&\cosh\beta&0&0\\
0&0&1&0\\
0&0&0&1
\end{bmatrix}
\end{align*}
The columns in the above matrices are labelled $(t,x,y,z)$ from left to right and likewise
for the rows. As we will latter be forming products of these matrices with the curvatures it
is sufficient to compute these matrices as if we were working in flat spacetime. Thus to
leading order in the lattice spacing we find\footnote{For the rotations we use standard
Euclidian trigonometry, for the boost we use the definition $n^\mu_i-n^\nu_j =
-K^\mu{}_\nu(x^\nu_i-x^\nu_j)$ where $n^\mu_a$ is the future pointing unit normal to the
Cauchy surface at the point $a$.}
\begin{gather*}
\cos\alpha = 1 + \BigO{L^3}\>,\quad \sin\alpha = \frac{d\Lxx}{dz} + \BigO{L^2}\\[5pt]
\cosh\beta = 1 + \BigO{L^3}\>,\quad \sinh\beta = -K_{xx}\Lxx + \BigO{L^2}
\end{gather*}
and thus
\bgroup
\def\P{\displaystyle K_{xx}\Lxx}
\def\Q{\displaystyle \frac{d\Lxx}{dz}}
\begin{equation*}
(U)_{10} = (B)_{10}(R)_{10} =
\begin{bmatrix}
1&-\P&0&0\\[5pt]
-\P&1&0&\Q\\[10pt]
0&0&1&0\\[5pt]
0&-\Q&0&1
\end{bmatrix} + \BigO{L^2}
\end{equation*}
\egroup
In a similar manner we find
\vspace{2pt}
\bgroup
\def\P{\displaystyle K_{yy}\Lyy}
\def\Q{\displaystyle \frac{d\Lyy}{dz}}
\begin{equation*}
(U)_{20} = (B)_{20}(R)_{20} =
\begin{bmatrix}
1&0&-\P&0\\[5pt]
0&1&0&0\\[10pt]
-\P&0&1&\Q\\[5pt]
0&0&-\Q&1
\end{bmatrix} + \BigO{L^2}
\end{equation*}
\egroup
\vspace{5pt}
\bgroup
\def\P{\displaystyle K_{xx}\Lxx}
\def\Q{\displaystyle \frac{d\Lxx}{dz}}
\begin{equation*}
(U)_{30} = (B)_{30}(R)_{30} =
\begin{bmatrix}
1&\P&0&0\\[5pt]
\P&1&0&-\Q\\[10pt]
0&0&1&0\\[5pt]
0&\Q&0&1
\end{bmatrix} + \BigO{L^2}
\end{equation*}
\egroup
\vspace{7pt}
\bgroup
\def\P{\displaystyle K_{yy}\Lyy}
\def\Q{\displaystyle \frac{d\Lyy}{dz}}
\begin{equation*}
(U)_{40} = (B)_{40}(R)_{40} =
\begin{bmatrix}
1&0&\P&0\\[5pt]
0&1&0&0\\[10pt]
\P&0&1&-\Q\\[5pt]
0&0&\Q&1
\end{bmatrix} + \BigO{L^2}
\end{equation*}
\egroup
For the remaining two matrices, $(U)_{50}$ and $(U)_{60}$, the job is quite simple, these
matrices are built solely on boosts. This leads to
\vspace{2pt}
\bgroup
\def\P{\displaystyle K_{zz}\Lzz}
\begin{equation*}
(U)_{50} = (B)_{50}(R)_{50} =
\begin{bmatrix}
1&0&0&-\P\\[5pt]
0&1&0&0\\[10pt]
0&0&1&0\\[5pt]
-\P&0&0&1
\end{bmatrix} + \BigO{L^2}
\end{equation*}
\egroup
\vspace{7pt}
\bgroup
\def\P{\displaystyle K_{zz}\Lzz}
\begin{equation*}
(U)_{60} = (B)_{60}(R)_{60} =
\begin{bmatrix}
1&0&0&\P\\[5pt]
0&1&0&0\\[10pt]
0&0&1&0\\[5pt]
\P&0&0&1
\end{bmatrix} + \BigO{L^2}
\end{equation*}
\egroup

Returning now to the construction of $(v)_{10}$, we have
\bgroup
\def\P{\displaystyle K_{xx}\Lxx}
\def\Q{\displaystyle \frac{d\Lxx}{dz}}
\begin{align*}
(v^\mu)_{10} &= (U^\mu{}_\nu)_{10} (v^\nu)_{11}\\
             &= (v^\mu)_{11}
              + \left[-\P v^x ,-\P v^t+\Q v^z,0,-\Q v^x\right]^\mu_{11}\\
\intertext{and}
(v^\mu)_{30} &= (U^\mu{}_\nu)_{30} (v^\nu)_{33}\\
             &= (v^\mu)_{33}
              + \left[\P v^x,\P v^t-\Q v^z,0,\Q v^x\right]^\mu_{33}
\end{align*}
\egroup
We are now in a position to finally compute $(v^t_{,x})_{00}$, to wit
\begin{align*}
(v^t_{,x})_{00} &= \frac{(v^t)_{10}-(v^t)_{30}}{2\Lxx} + \BigO{L^a}\\
                &= \frac{(v^t)_{11}-(v^t)_{33}}{2\Lxx} 
                 - K_{xx}\frac{(v^x)_{11}+(v^x)_{33}}{2} + \BigO{L^a}
\end{align*}
Here we have written the truncation errors as $\BigO{L^a}$ with $a>0$ for it is not clear, at
this level of analysis, what the exact nature of this term is (save that it vanishes as
$L\rightarrow0$). Since our spacetime is spherically symmetric we have
\begin{equation*}
(v)_{00} = (v)_{11} = (v)_{22} = (v)_{33} = (v)_{44} 
\end{equation*}
and thus
\begin{equation*}
(v^t_{,x})_{00} = -K_{xx}(v^x)_{00} + \BigO{L^a}
\end{equation*}
Similar calculations can be used to compute all of the spatial derivatives of $v^\mu$ at the
central vertex.

We can now return to the principle objective of this section -- to compute the various partial
derivatives of the curvatures. We proceed exactly as above but with a minor change in that we
will no longer carry the truncation errors within the calculations. Thus we have
\begin{equation*}
(\rmanb)_{i0} = (U_{\mu}{}^{\tau})_{i0}(U_{\alpha}{}^{\rho})_{i0}
                (U_{\nu}{}^{\delta})_{i0}(U_{\beta}{}^{\lambda})_{i0}(\rtrdl)_{ii}
\end{equation*}
for $i=1,2,3,4,5,6$ and $(U_{\mu}{}^{\nu})_{i0} =
g_{\mu\rho}g^{\nu\tau}(U^{\rho}{}_{\tau})_{i0}$ with $g_{\mu\nu} = \diag(-1,1,1,1)$. And, as
before,
\begin{gather*}
(\rmanb)_{00} = (\rmanb)_{11} = (\rmanb)_{22} = (\rmanb)_{33} = (\rmanb)_{44}
\end{gather*}
due to spherical symmetry.

Using the above expressions for the $(U)_{i0}$ and the following finite difference
approximations
\begin{align*}
(\dRtyxyx)_{00} &= \frac{(\rtyxy)_{10}-(\rtyxy)_{30}}{2\Lxx}\\[5pt]
(\dRtxxyy)_{00} &= \frac{(\rtxxy)_{20}-(\rtxxy)_{40}}{2\Lyy}\\[5pt]
(\dRtzxzx)_{00} &= \frac{(\rtzxz)_{10}-(\rtzxz)_{30}}{2\Lxx}\\[5pt]
(\dRtxxzz)_{00} &= \frac{(\rtxxz)_{50}-(\rtxxz)_{60}}{2\Lzz}
\end{align*}
we find that
\begin{align}
\dRtyxyx &= \phantom{-}\kxx\left(\rxyxy+\rtyty\right) 
            + \frac{1}{\Lxx}\DLxxDz \rtyyz\label{eqn:DerivRa}\\[5pt]
\dRtxxyy &= -\kyy\left(\rxyxy+\rtxtx\right) 
            - \frac{1}{\Lyy}\DLyyDz \rtxxz\label{eqn:DerivRb}\\[5pt]
\dRtzxzx &= \phantom{-}\kxx\left(\rxzxz+\rtztz\right) 
            + \frac{1}{\Lxx}\DLxxDz \rtxxz\label{eqn:DerivRc}\\[5pt]
\dRtxxzz &= -\kzz\left(\rxzxz+\rtxtx\right)\label{eqn:DerivRd}
\end{align}
We have dropped the $00$ subscript as we no longer need to distinguish between the
neighbouring frames. By spherical symmetry we have
\begin{gather*}
\Lxx = \Lyy\>,\quad \kxx=\kyy\>,\quad \rtxxz=\rtyyz\>,\quad \rxzxz=\ryzyz
\end{gather*}
while from the vacuum Einstein equations we have
\begin{align*}
0 &= \rtz = -\rtxxz - \rtyyz\\
0 &= \rxx =  \rxyxy + \rxzxz - \rtxtx\\
0 &= \ryy =  \rxyxy + \ryzyz - \rtyty\\
0 &= \rzz =  \rxzxz + \ryzyz - \rtztz
\end{align*}
Combining the last few equations leads to
\begin{gather*}
\rtxtx = \rtyty = \rxyxy+\rxzxz\\
\rtxxz = \rtyyz = 0\>,\quad  \rtztz = 2\rxzxz
\end{gather*}
Substituting these into the above equations (\ref{eqn:DerivRa}--\ref{eqn:DerivRd}) and
subsequently into the previous expressions for the Bianchi identities
(\ref{eqn:RiemEvolA},\ref{eqn:RiemEvolD}) leads to the following pair of equations
\begin{align*}
\DrxDt &= 2\kxx\left( 2\rxyxy  + \rxzxz\right)\\[5pt]
\DrzDt &= 3\kxx\rxzxz + \kzz\left(\rxyxy + 2\rxzxz\right)
\end{align*}
Our job is almost complete, but we still have two tasks ahead of us i) to introduce a lapse
function and ii) to account for the limited time interval over which a single Riemann normal
frame can be used. The first task is rather easy, we simply make the coordinate substitution
$t\rightarrow Nt$ leading to
\begin{align}
\DrxDt &= 2N\kxx\left( 2\rxyxy  + \rxzxz\right)\label{eqn:EvolveA}\\[5pt]
\DrzDt &= 3N\kxx\rxzxz + N\kzz\left(\rxyxy + 2\rxzxz\right)\label{eqn:EvolveB}
\end{align}
and where we now have $(\gmn)_o=\diag(-N^2,1,1,1)$. The lapse $N$ can be freely chosen at each
vertex of the lattice (but subject to the obvious constraint that $N>0$). The second task is a
bit more involved. We know that each Riemann normal frame is limited in both space and time.
Thus no single Riemann normal frame can be used to track the evolution for an extended period
of time. We will have no choice but to jump periodically to a new frame. This can be elegantly
handled in the moving frame formalism. Thus our task reduces to finding a new set of evolution
equations for the frame components $\Rxyxy$ and $\Rxzxz$ based on the equations given above
for $\rxyxy$ and $\rxzxz$.

Let $e^\mu{}_a$, $a=t,x,y,z$ be an orthonormal tetrad\footnote{This tetrad is identical to
that used in section (\ref{sec:SphericalSpacetime}), the change of notation introduced here is
simply to avoid unwanted clutter in the following equations.}, tied to the worldline of the
central vertex and aligned to the coordinate axes. Thus we have $e^\mu{}_t$ as the future
pointing tangent vector to the worldline while $e^\mu{}_z$ points along the $z$-axis. Then
\begin{align*}
\DRxDt =& \frac{d}{dt}\left(\rmanb \emx \eay \enx \eby \right)\\[5pt]
\DRzDt =& \frac{d}{dt}\left(\rmanb \emx \eaz \enx \ebz \right)
\end{align*}
Since our spacetime is spherically symmetric it is not hard to see that the tetrads of two
consecutive cells (on the vertex worldline) are related by a boost in the $t-z$ plane
(arising from gradients in the lapse function). A simple calculation shows that
\begin{gather*}
\DemxDt = 0\>,\quad \DemyDt = 0\>,\quad
\DemtDt = \dNdz \emz\>,\quad \DemzDt = \dNdz \emt
\end{gather*}
which when combined with the above leads to
\begin{align}
\DRxDt =& \left(\DrmanbDt\right) \emx \eay \enx \eby\\[5pt]
\DRzDt =& \left(\DrmanbDt\right) \emx \eaz \enx \ebz 
         -2 \frac{N_{,z}}{N} \rmanb \emt \eax \enx \ebz
\end{align}
In our frame we have chosen $\left(\gmn\right)_{o}=\diag(-N^2,1,1,1)$,
$e^\mu{}_{a}=\delta^\mu_a$ for $a=x,y,z$ and $e^\mu{}_{t}=1/N$, thus we see that the last term
in the previous equation is proportional to $\Rtxxz$. But for the Schwarzschild spacetime we
know that $\Rtxxz=0$ and thus we have
\begin{align}
\DRxDt =& \left(\DrmanbDt\right) \emx \eay \enx \eby\\[5pt]
\DRzDt =& \left(\DrmanbDt\right) \emx \eaz \enx \ebz
\end{align}
which, when combined with (\ref{eqn:EvolveA},\ref{eqn:EvolveB}), leads immediately to the
evolution equations (\ref{eqn:EvolveRx},\ref{eqn:EvolveRz}) quoted in section
(\ref{sec:EvolveEqtns}).

\clearpage


\captionsetup{margin=0pt,font=small,labelfont=bf}

\def\Figure#1#2{%
\centerline{%
\includegraphics[width=#1\textwidth]{#2}}
\vskip0.5cm}

\def\FigPair#1#2{%
\centerline{%
\includegraphics[width=0.6\textwidth]{#1}\hfill%
\includegraphics[width=0.6\textwidth]{#2}}
\vskip0.5cm}

\def\FigPairV#1#2{%
\centerline{%
\includegraphics[width=0.95\textwidth]{#1}}
\vskip0.25cm
\centerline{%
\includegraphics[width=0.95\textwidth]{#2}}
\vskip0.5cm}

\def\FigQuad#1#2#3#4{%
\centerline{%
\includegraphics[width=0.6\textwidth]{#1}\hfill%
\includegraphics[width=0.6\textwidth]{#2}}%
\centerline{%
\includegraphics[width=0.6\textwidth]{#3}\hfill%
\includegraphics[width=0.6\textwidth]{#4}}
\vskip0.5cm}


\begin{figure}[t]
\Figure{1.0}{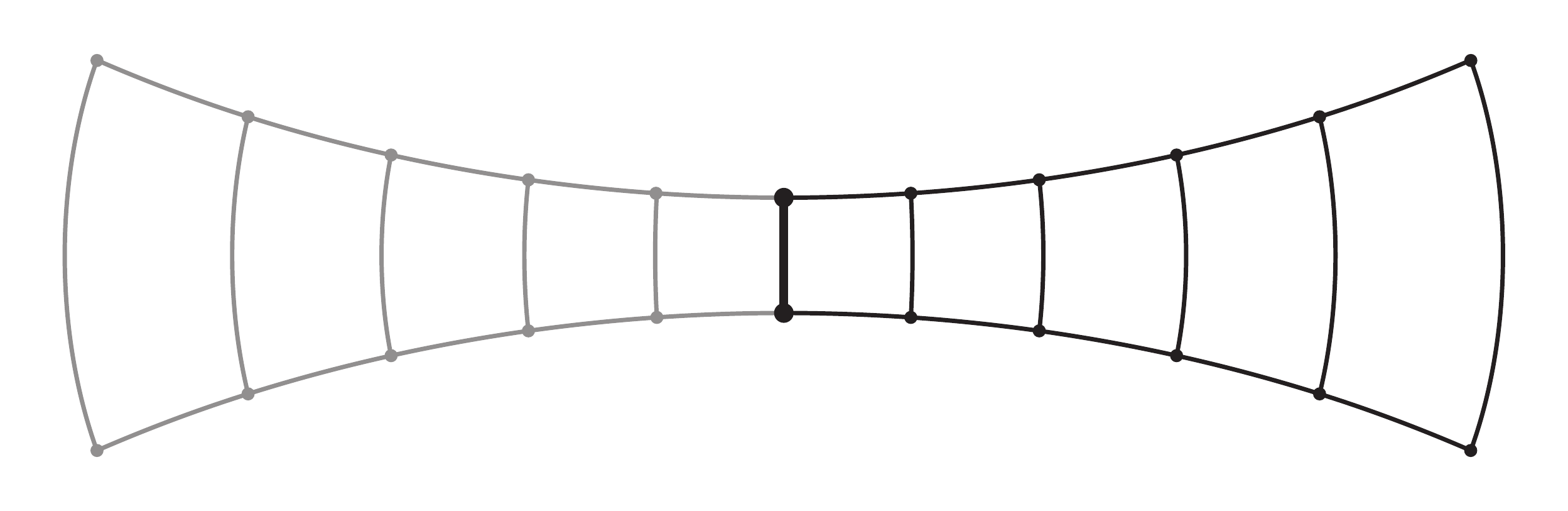}
\caption{\normalfont%
A simple lattice for a Schwarzschild spacetime. This consists of two identical halves joined
at the throat (denoted by the thick line). In our computer code we only store the right hand
half (plus a few nodes from the left half to ensure reflection symmetry at the throat).}
\label{fig:SchwarzLattice}
\end{figure}

\begin{figure}[t]
\Figure{0.85}{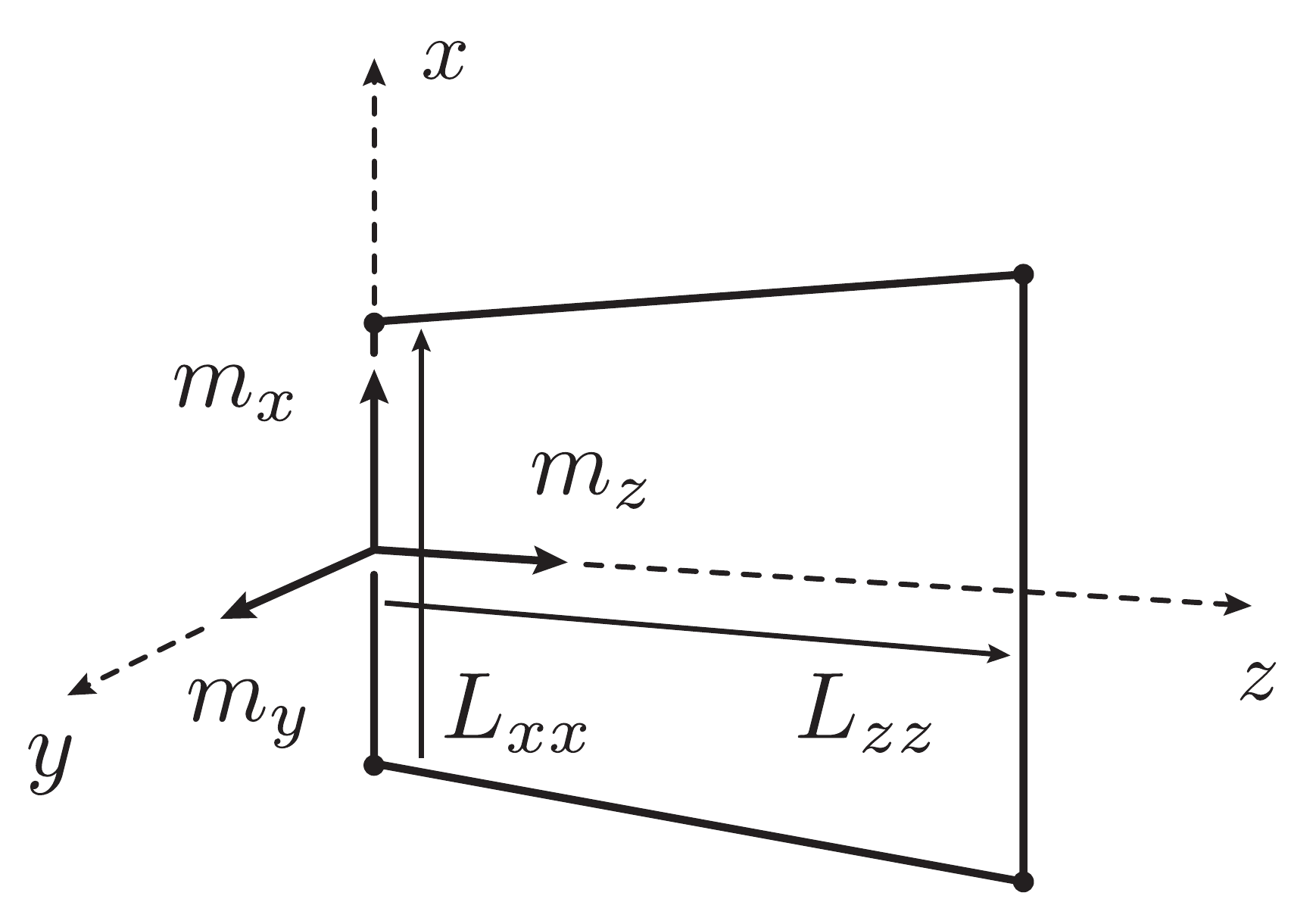}
\caption{\normalfont%
The local structure of the lattice. The $(x,y,z)$ are a set of coordinates local
to this set of legs. There is one such coordinate frame for each $\Lxx$ along the
lattice. These coordinates are never used in the computer code but help to define the metric
in the neighbourhood of $\Lxx$.}
\label{fig:LatticeLocal}
\end{figure}

\begin{figure}[t]
\Figure{0.85}{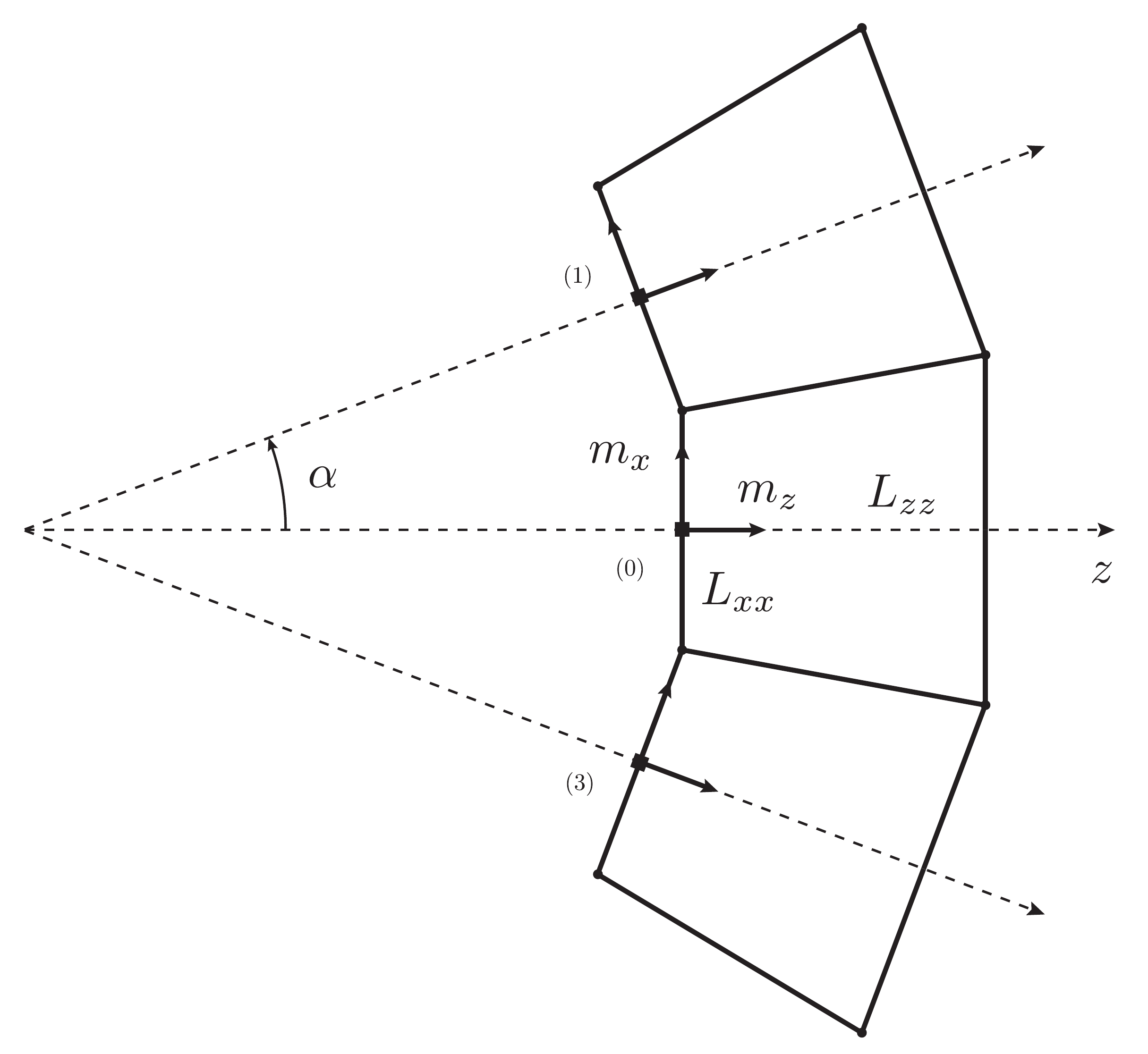}
\caption{\normalfont%
This shows an $xz$ section of the lattice obtained by cloning the original 2-dimensional
lattice. The data in the upper and lower cells are identical to that in the middle cell, this
follows from spherical symmetry. The small squares denote the central vertices of each
computational cell. The angle $\alpha$ can be computed using standard Euclidean trigonometry
as described in the text.}
\label{fig:LatticeCloned}
\end{figure}


\begin{figure}[t]
\FigPairV{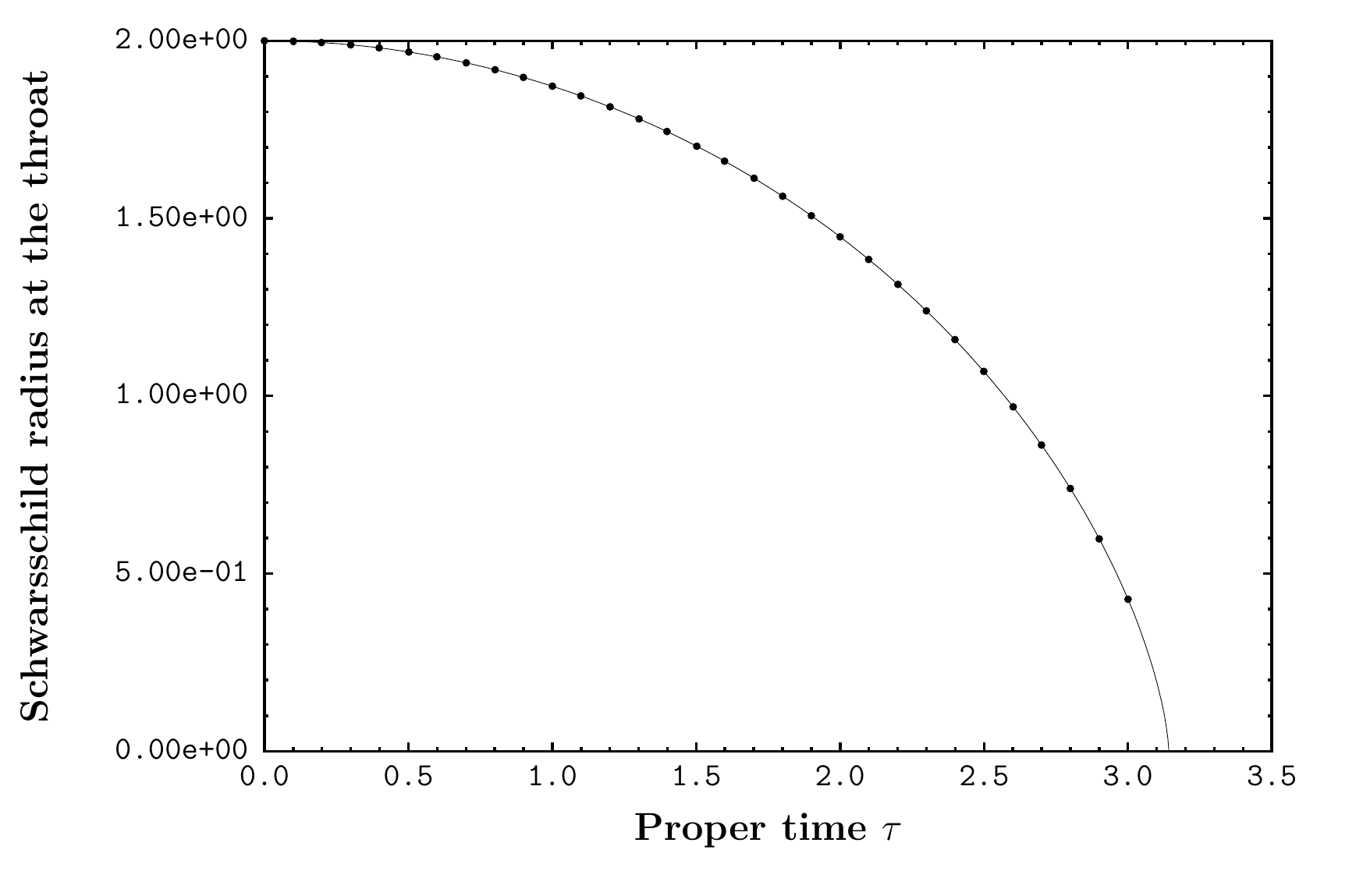}%
         {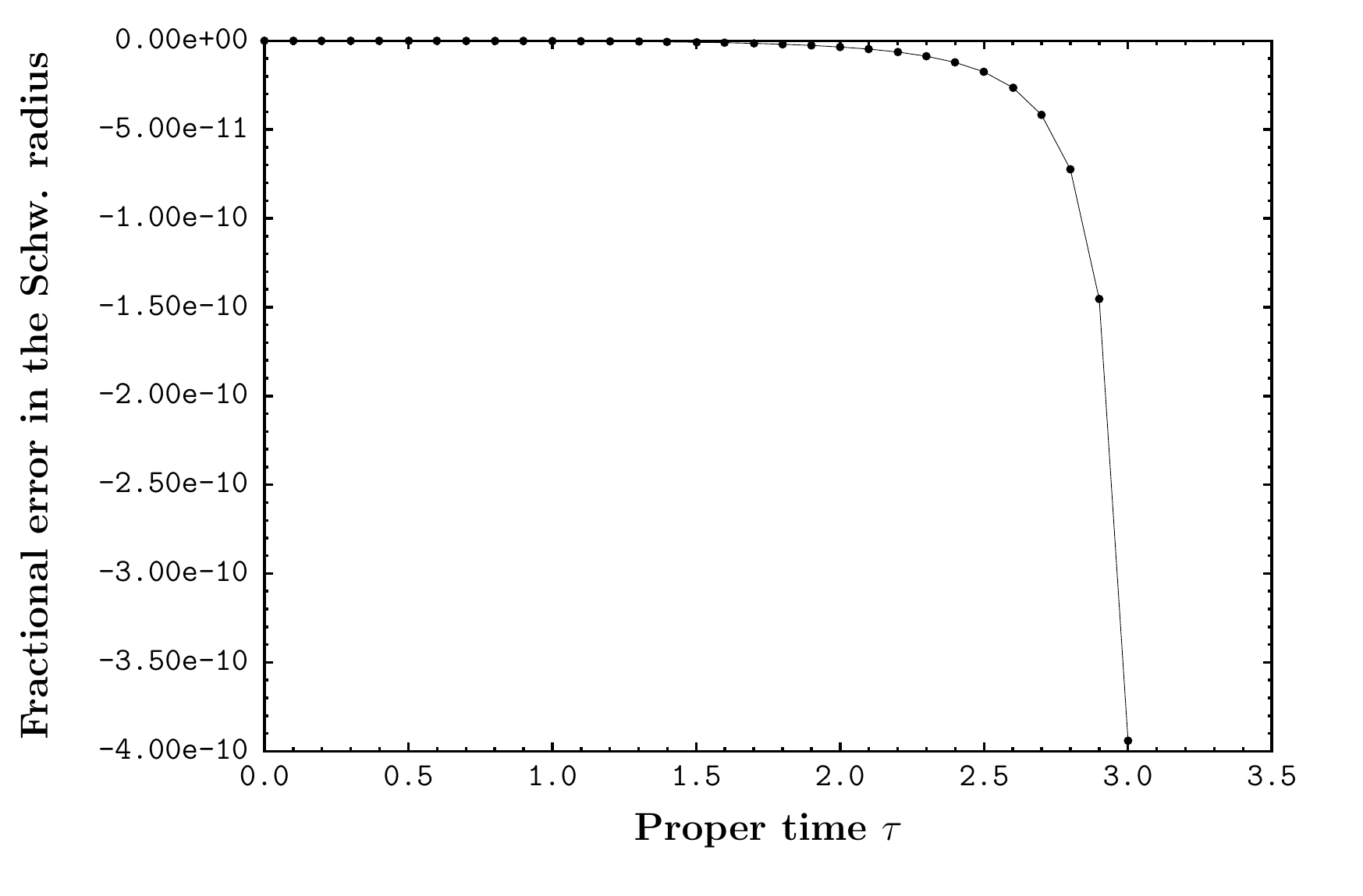}
\caption{\normalfont%
The Schwarzschild areal coordinate $r$ and the lattice $\Lxx$ at the throat are
related by $r(t) = 2m\Lxx(t)/\Lxx(0)$. In geodesic slicing $r(t)$ at the throat is described
by the parametric equations $r(t)=m(1+\cos\eta(t))$, $t(\eta) = m(\eta+\sin\eta)$. These
equations allow us to plot the exact evolution of $r(t)$ (the smooth curve) against
estimates from the lattice (solid points). The relative errors are seen to be very small
and are dominated by the truncation errors in the Runge-Kutta scheme.}
\label{fig:GeodesicLapse}
\end{figure}

\begin{figure}[t]
\FigQuad{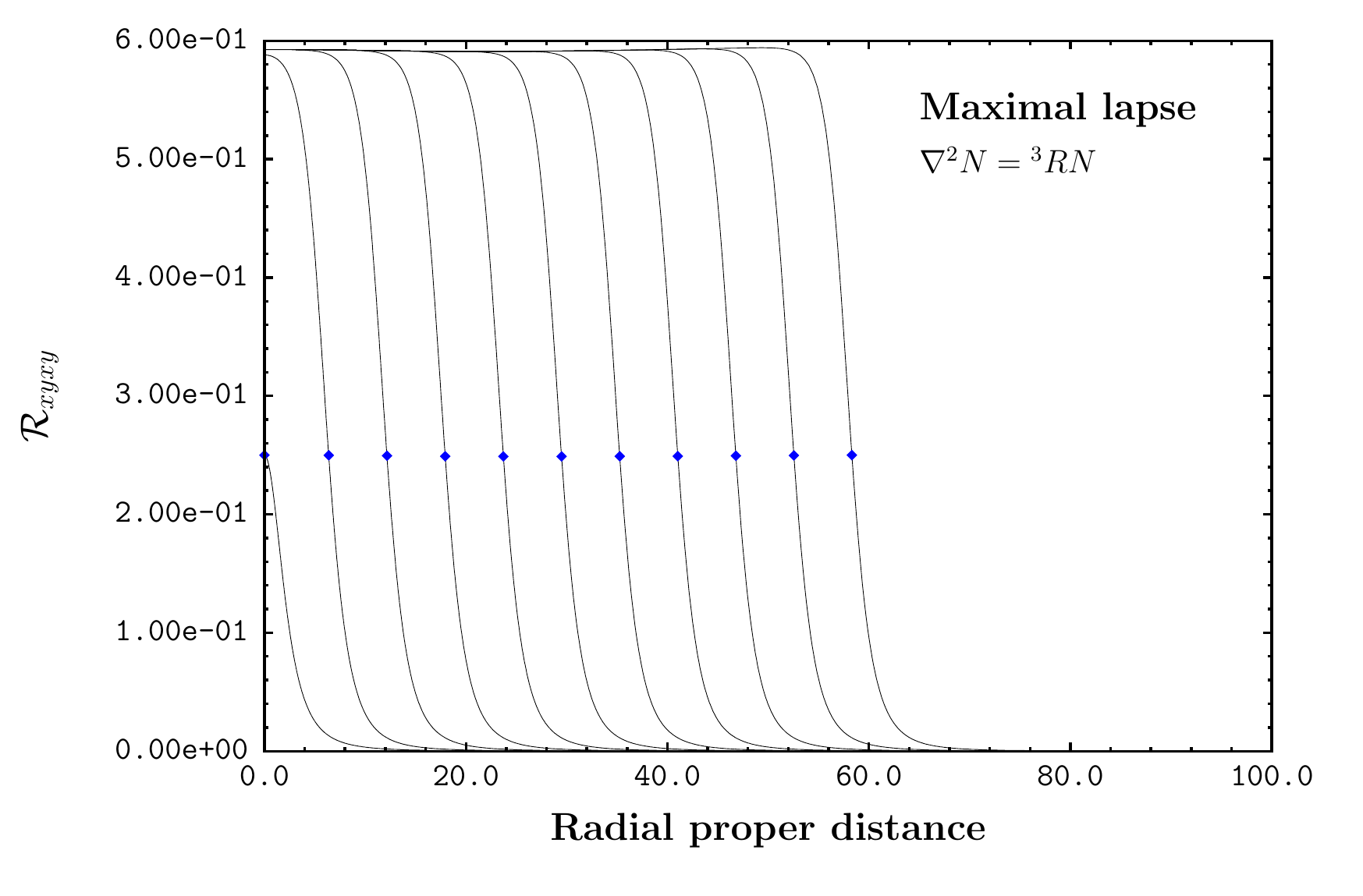}%
        {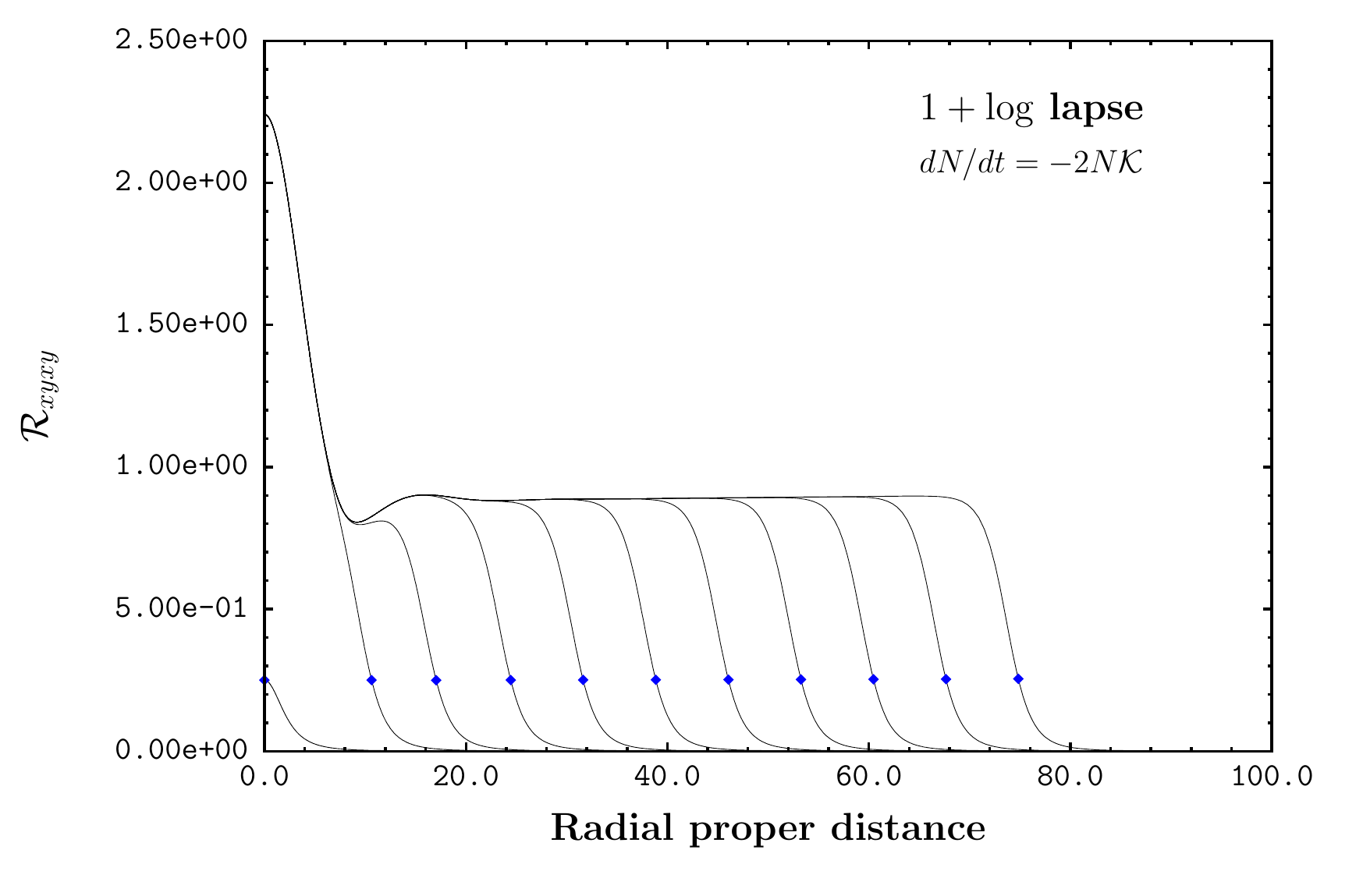}
        {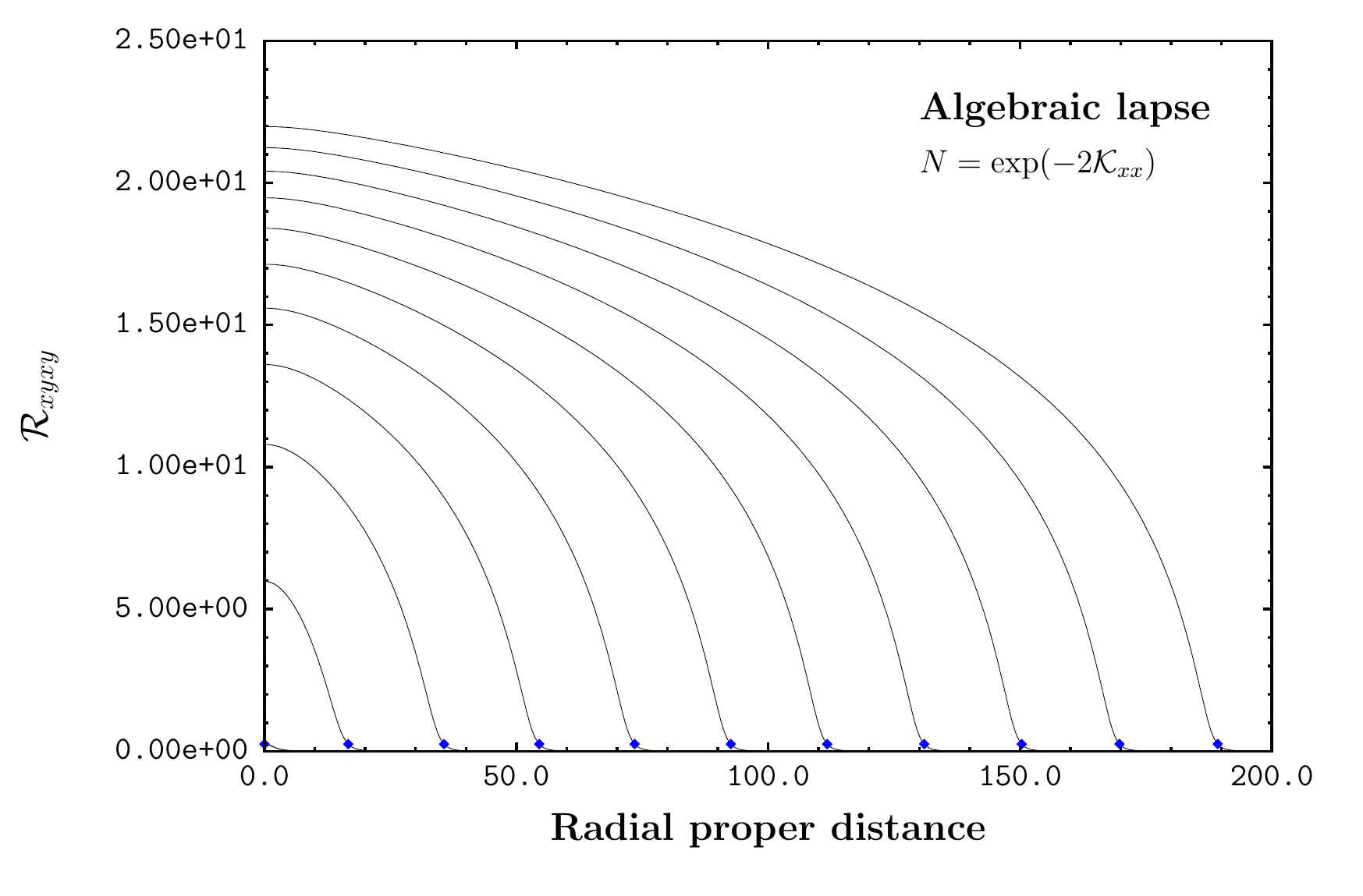}%
        {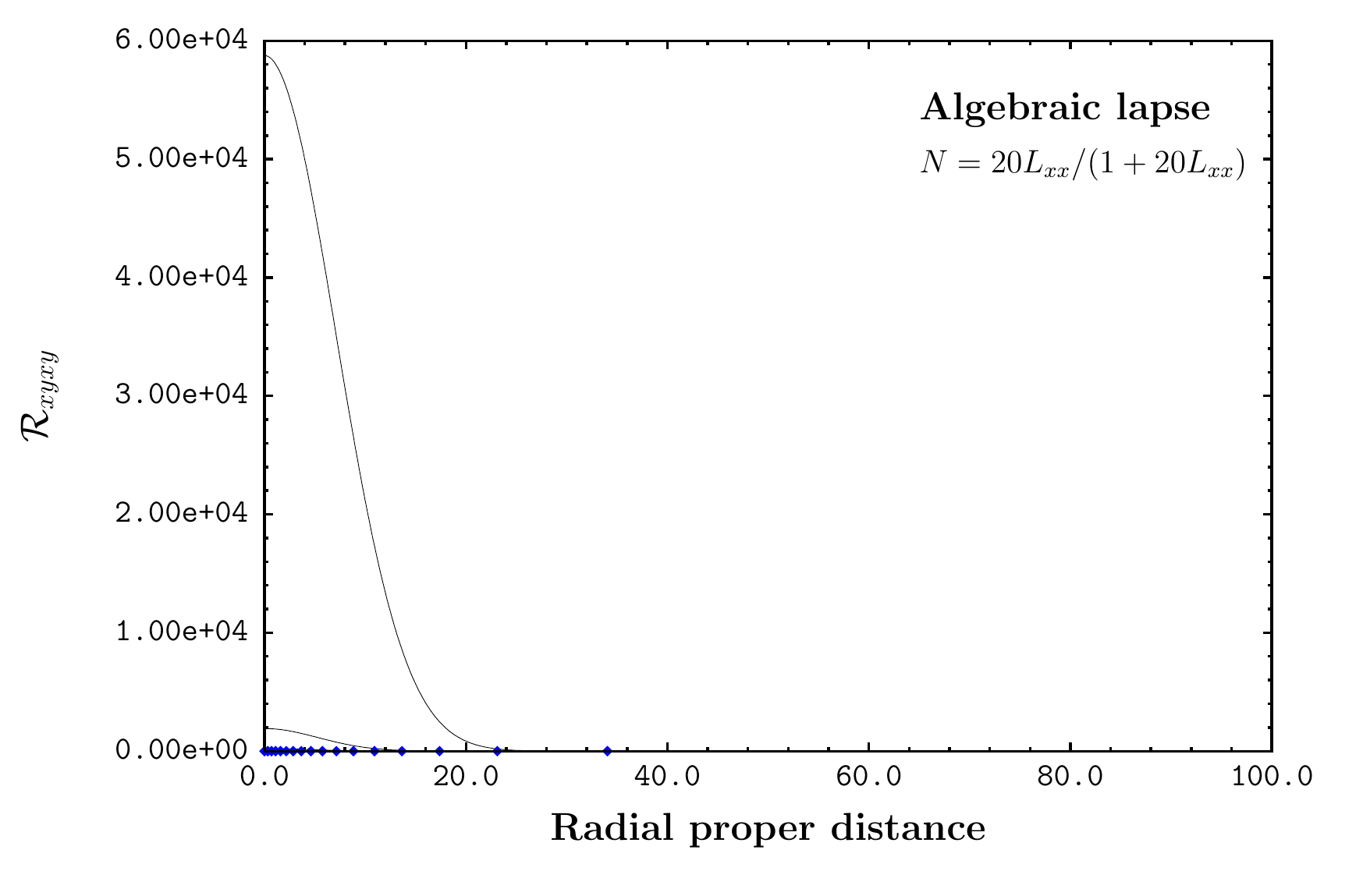}
\caption{\normalfont%
The curvature profiles for four choices of lapse function. Each figure shows the lapse
function for $t=0$ to $t=100$ in steps of 10. The small diamond on each curve represents the
location of the apparent horizon. There are only two curves visible in the algebraic slicing
$N=20\Lxx/(1+20\Lxx)$
due to the rapid rise in the curvatures as the slicing approaches the singularity.}
\label{fig:CurvatureProfiles}
\end{figure}

\begin{figure}[t]
\FigQuad{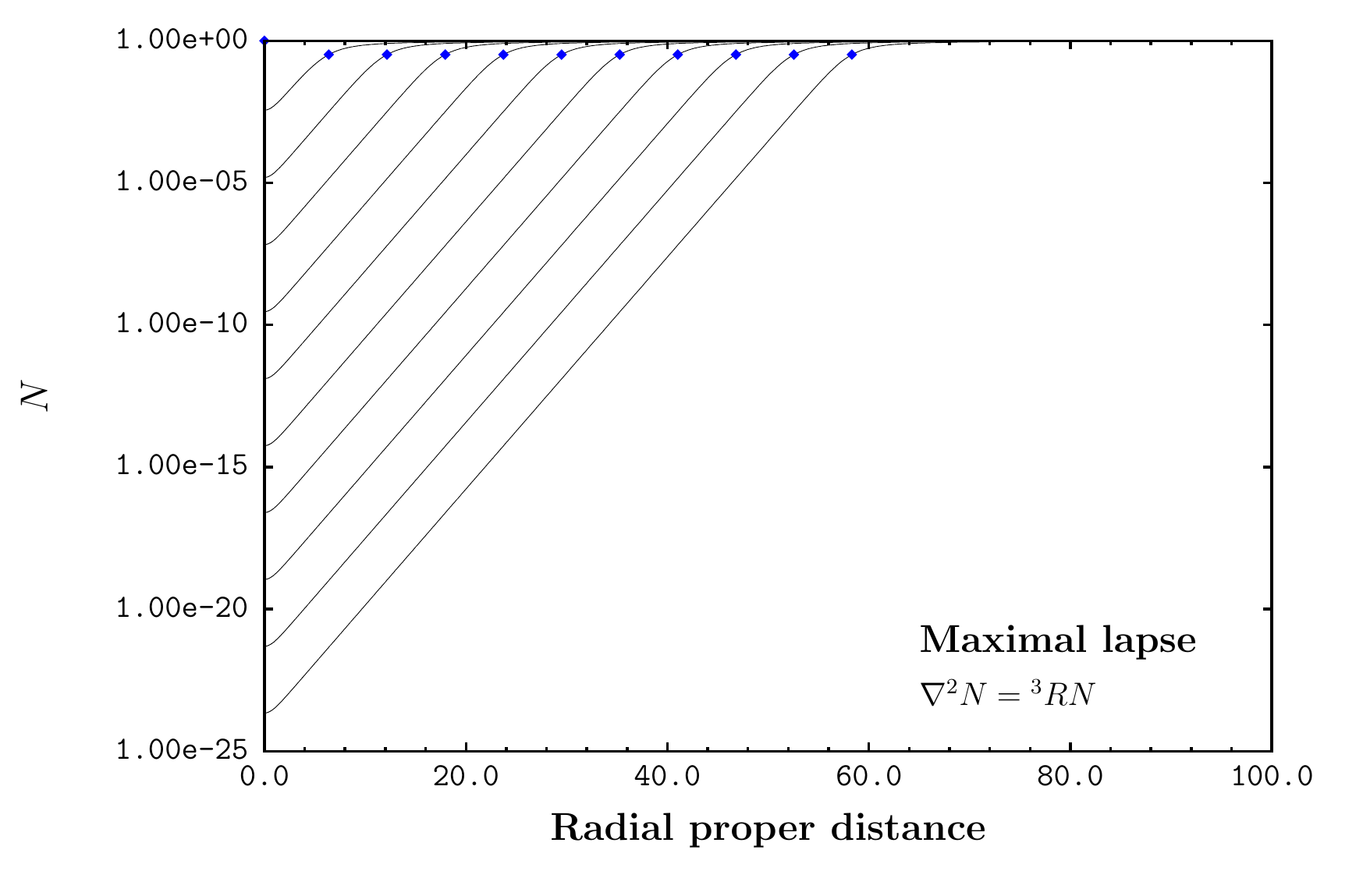}%
        {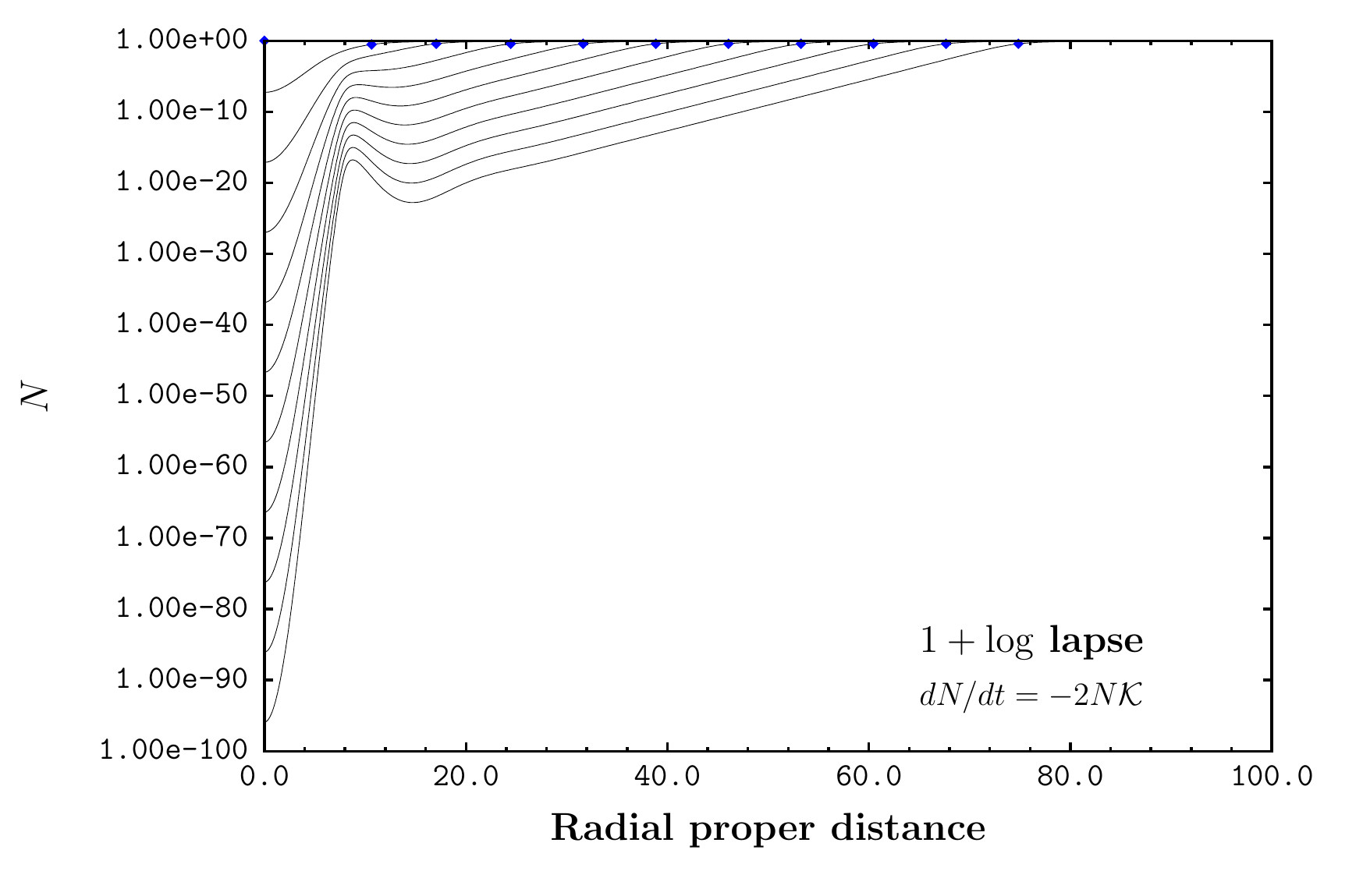}
        {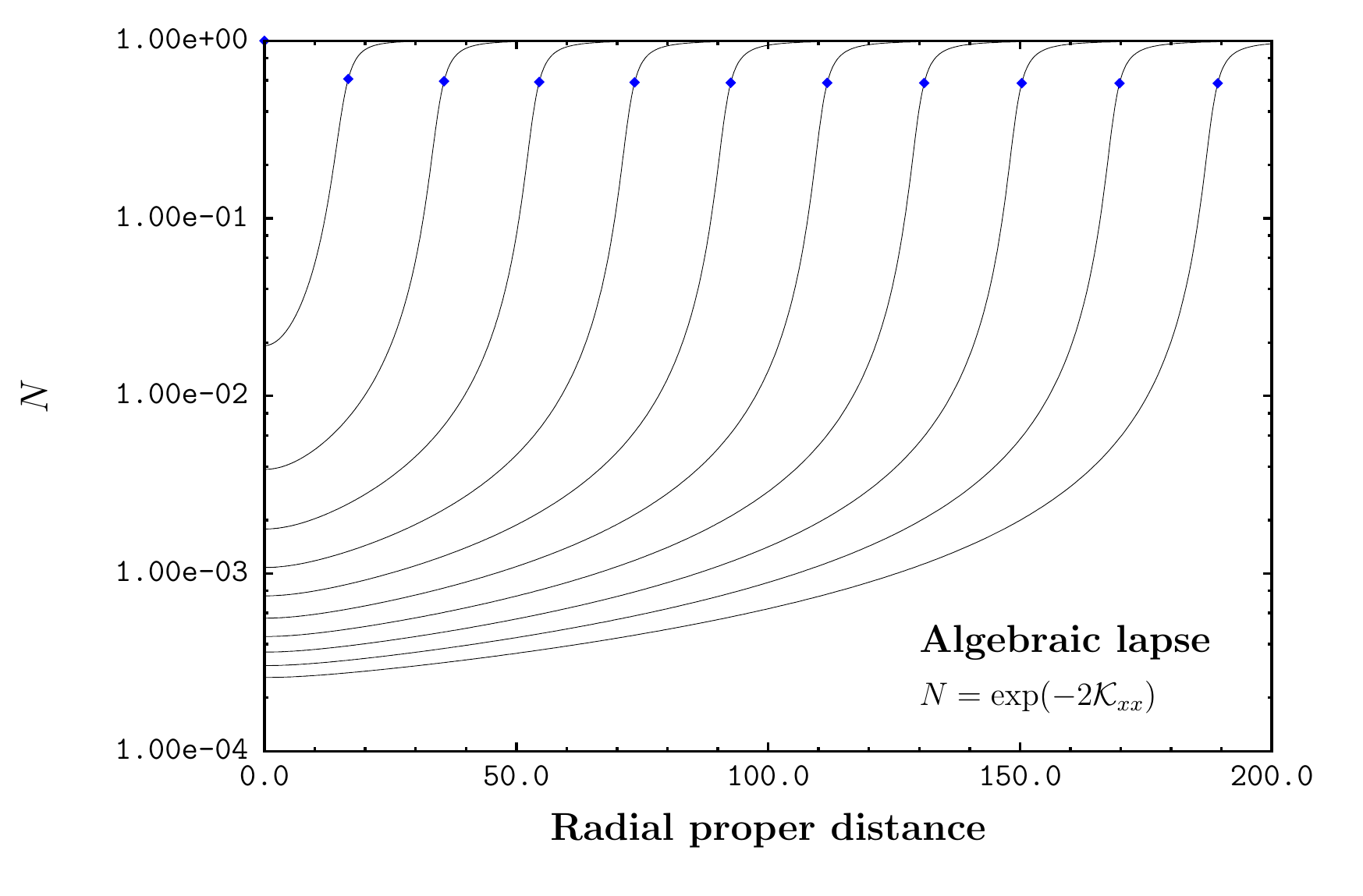}%
        {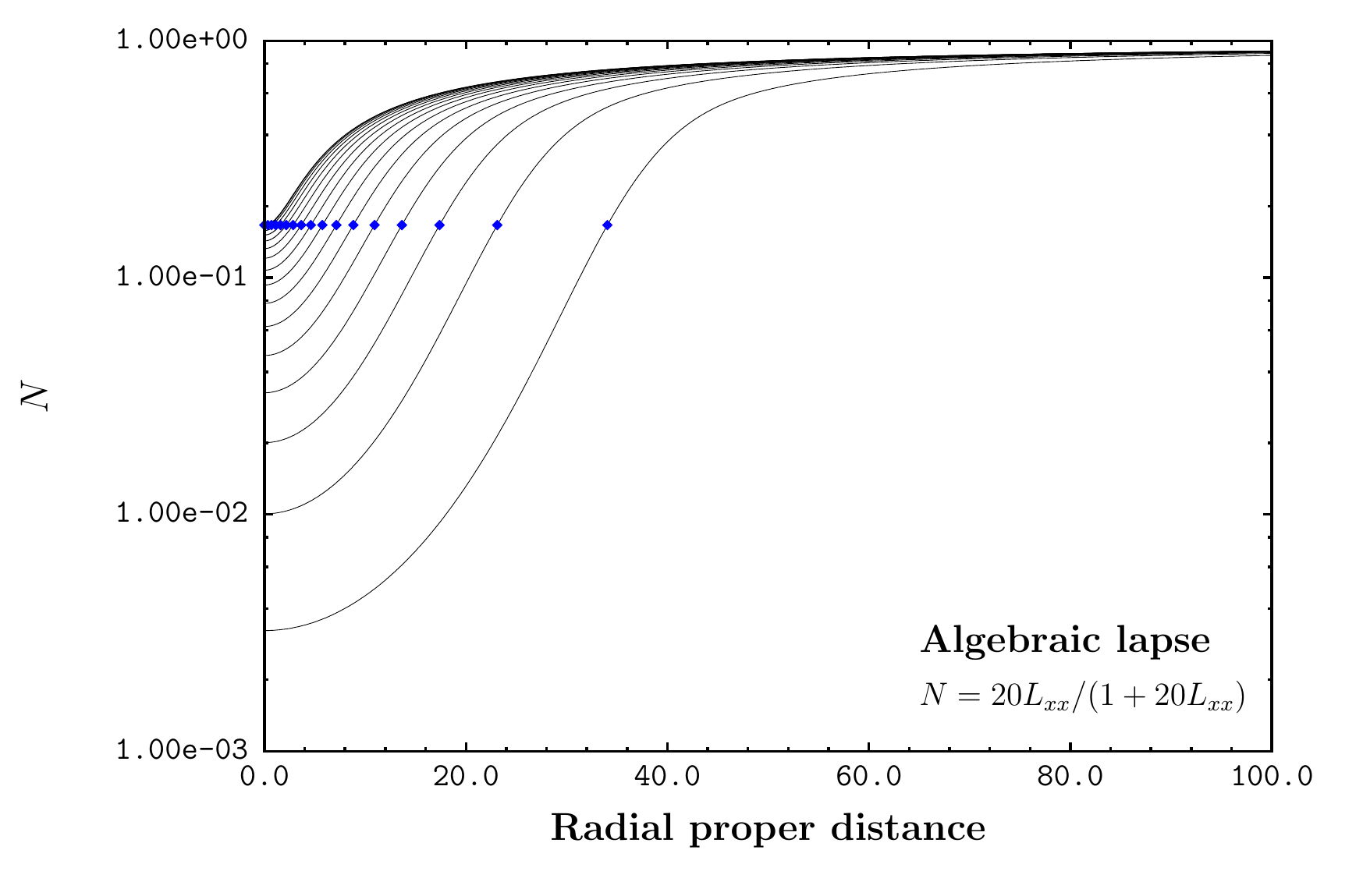}%
\caption{\normalfont%
As per figure (\ref{fig:CurvatureProfiles}) but this time we display the lapse profiles. Note
how quickly the lapse collapse at the throat in the $1+\log{}$ slicing. This would likely
cause serious underflow problems for $t\gtrsim 300$. Notice also the uniform spacing of the
curves along the logarithmic axis for the maximal lapse. This show that the lapse
collapses exponentially at the throat (a well known result for maximal slicing).}
\label{fig:LapseProfiles}
\end{figure}

\begin{figure}[t]
\FigQuad{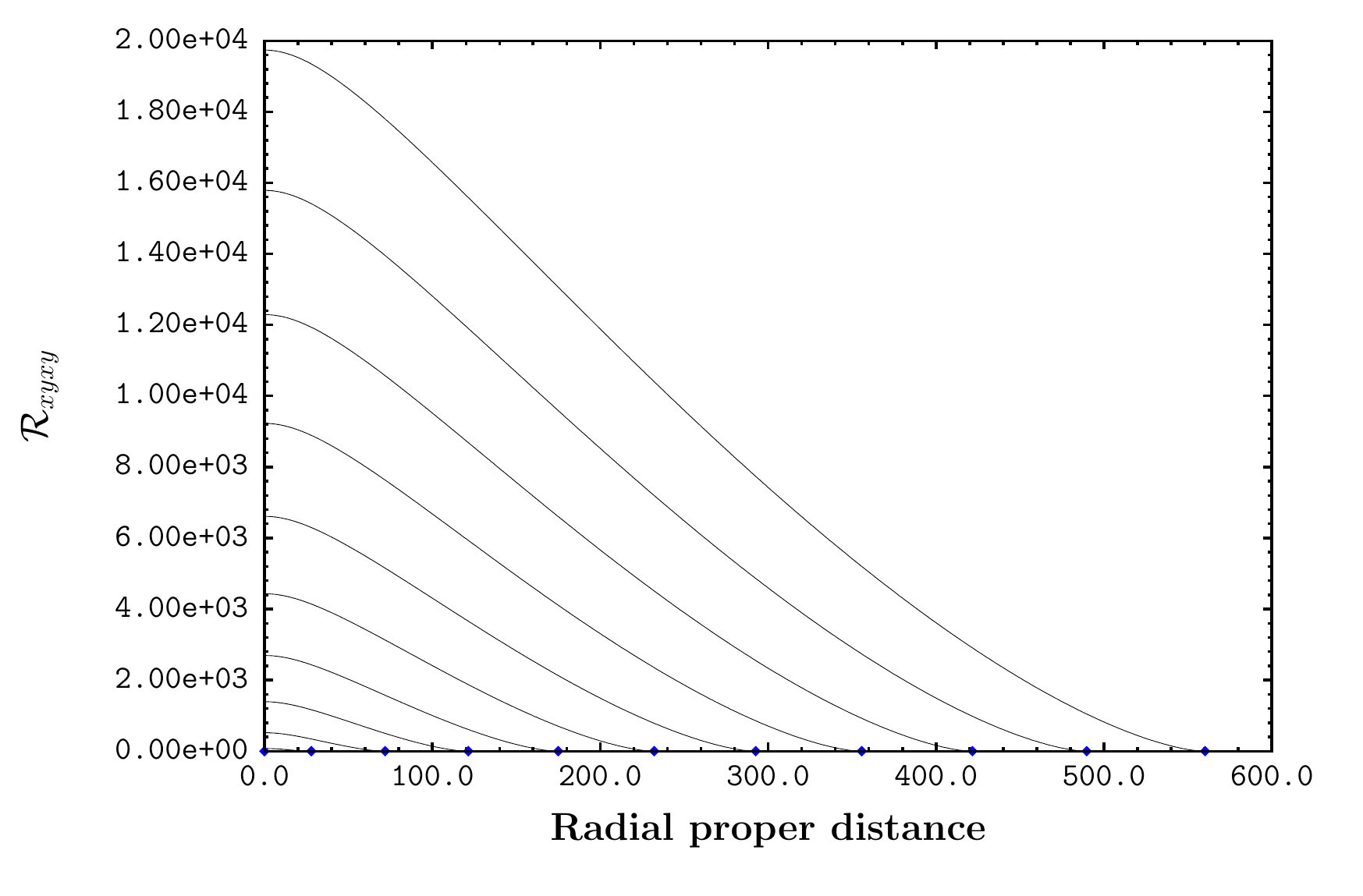}%
        {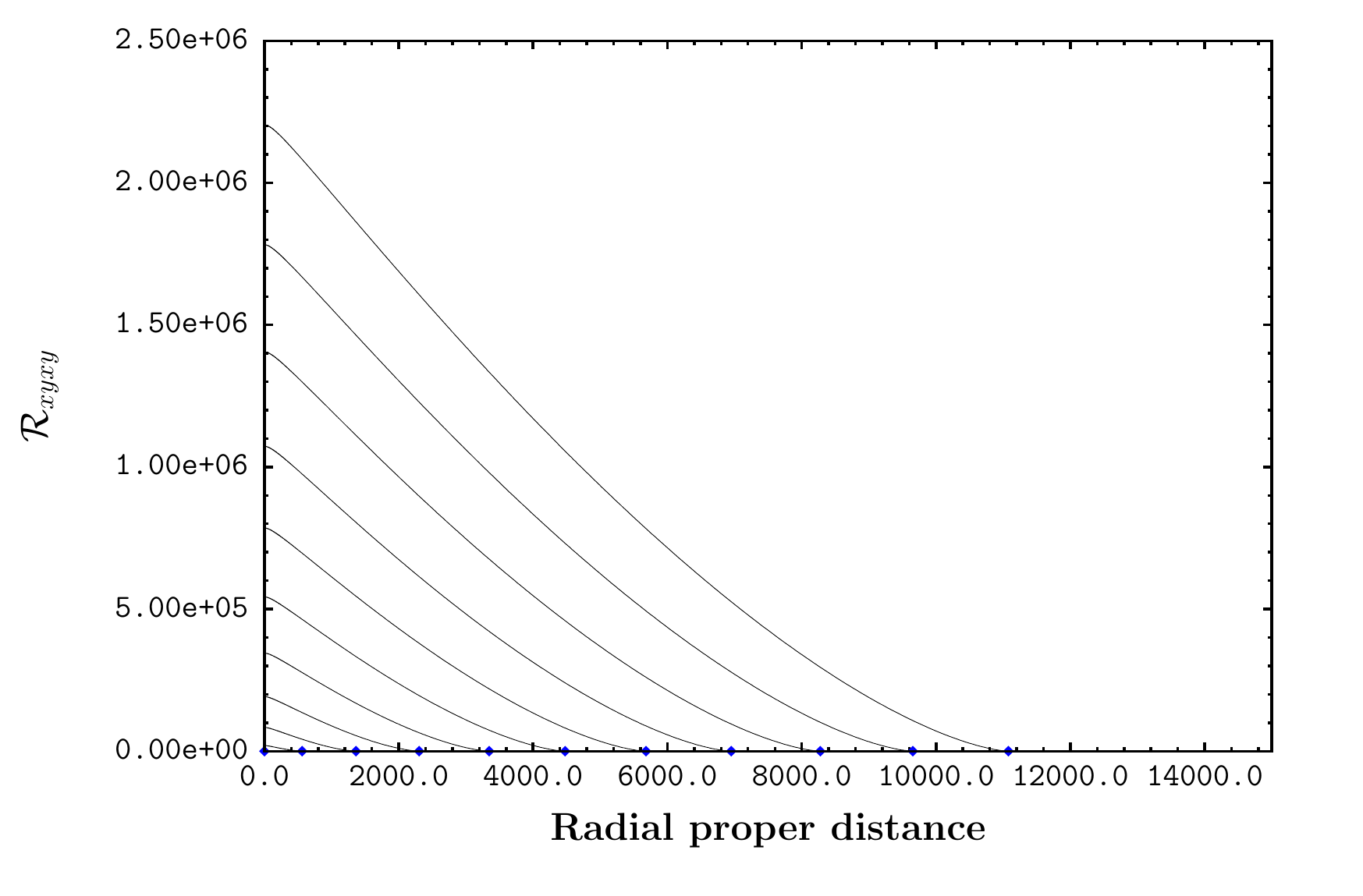}%
        {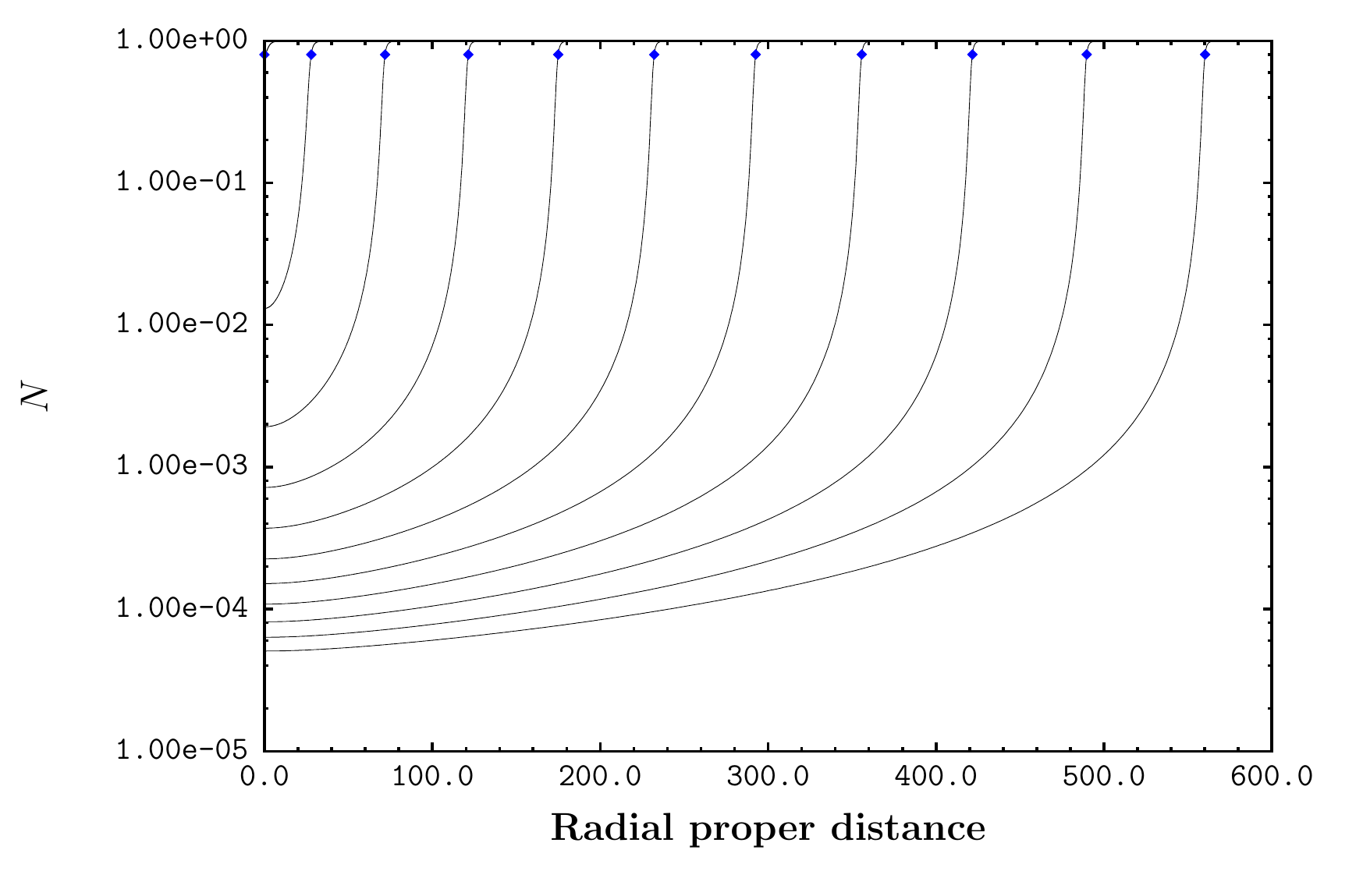}%
        {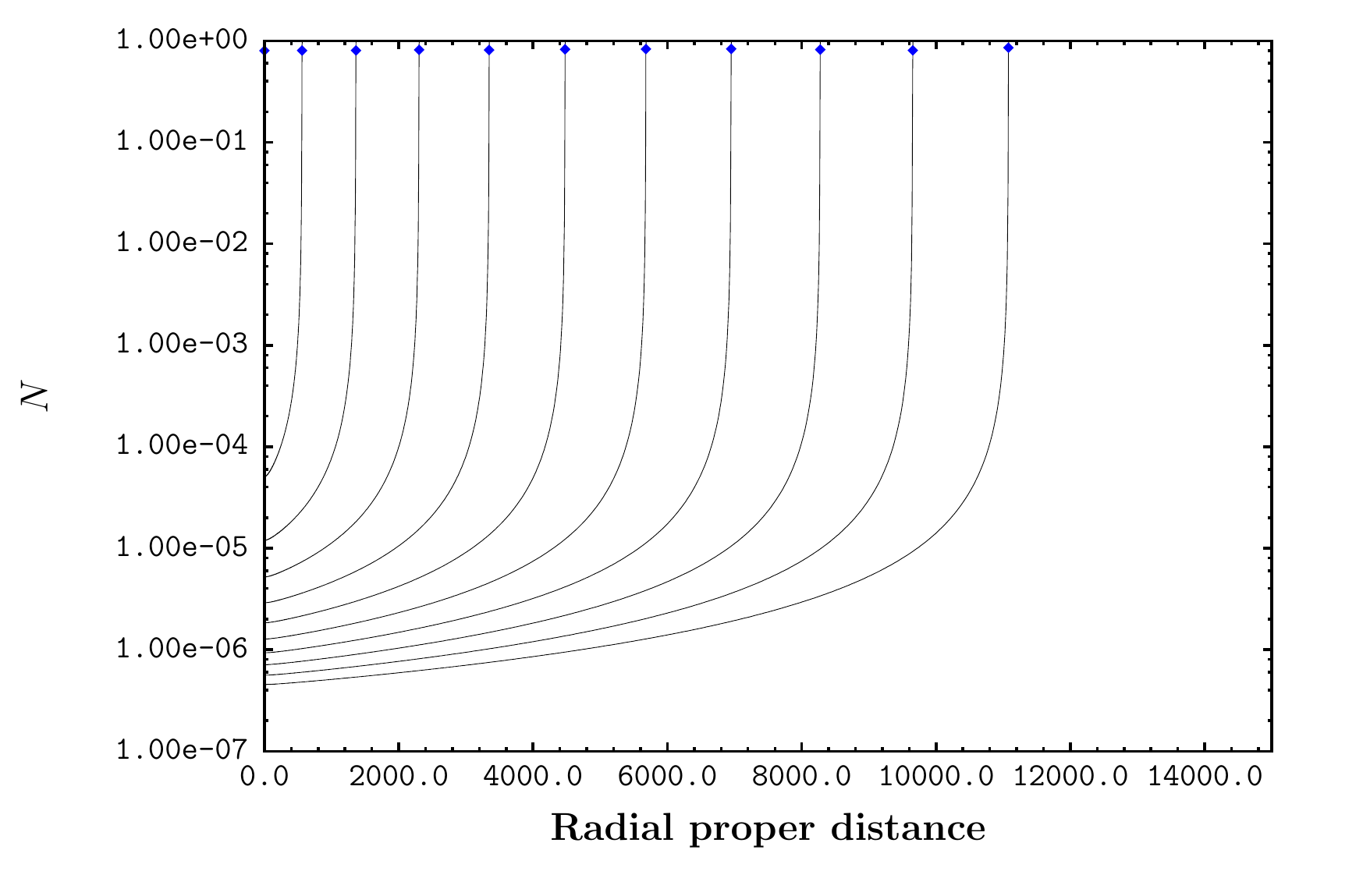}
\caption{\normalfont%
The Riemann curvature $\Rx$ and the lapse function $N$ for the algebraic slicing with
$N=1/(1+\Rx)$. The left pair of figures shows the evolution over the range $t=0$ to 100 in
steps of 10, while the right pair cover the range $t=0$ to $1000$ in steps of 100. This lapse
does not appear (on this time scale) to exhibit an exponential collapse at the throat. The
profiles for the algebraic slicings appear to propagate into the asymptotically flat regions
far more rapidly than any of the differential slicings. This may be explain why an artificial
viscosity was not needed for the algebraic slicings -- they carry away
any small numerical errors before they have chance to grow.}
\label{fig:algebraic}
\end{figure}

\begin{figure}[t]
\FigQuad{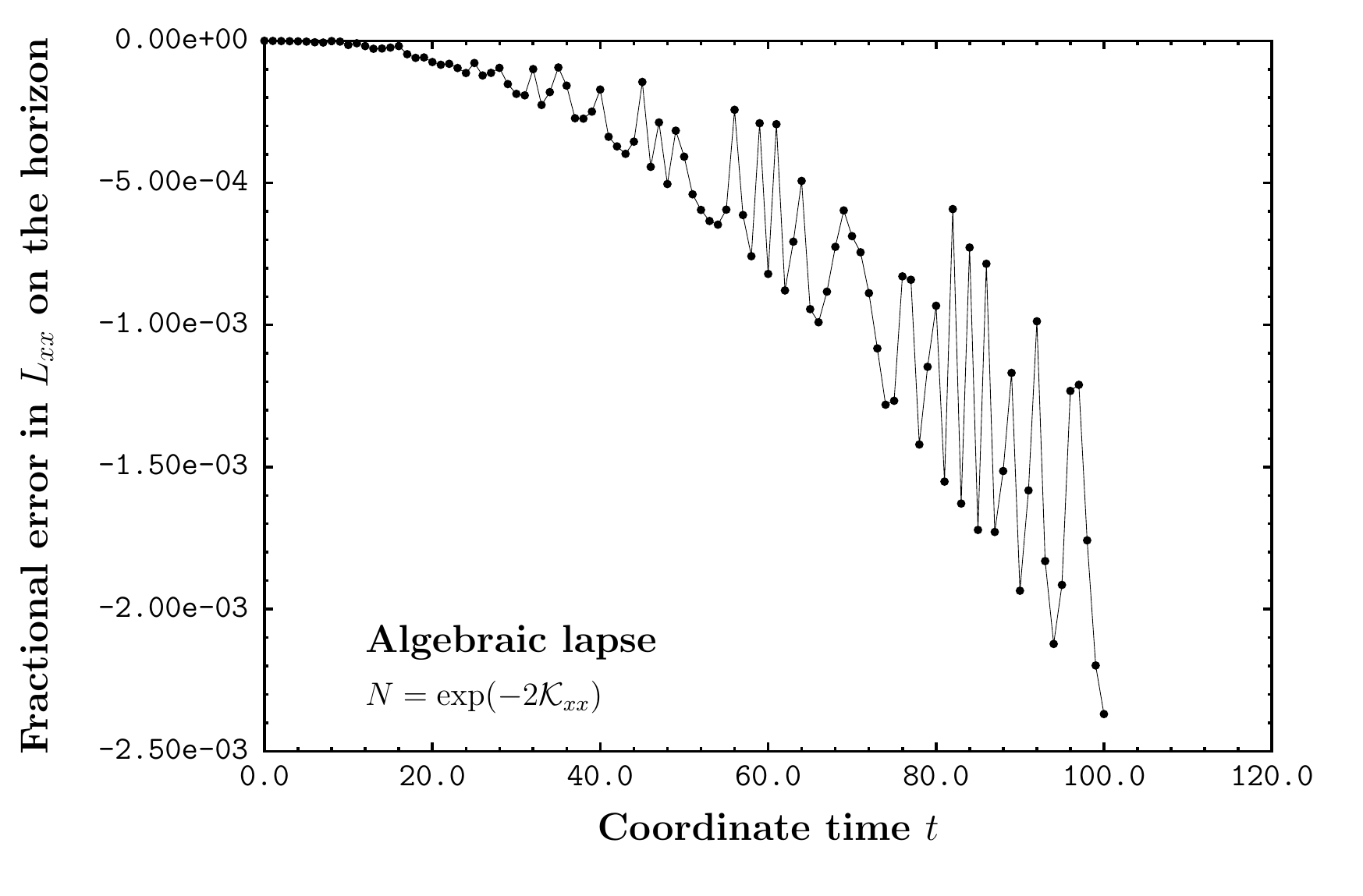}%
        {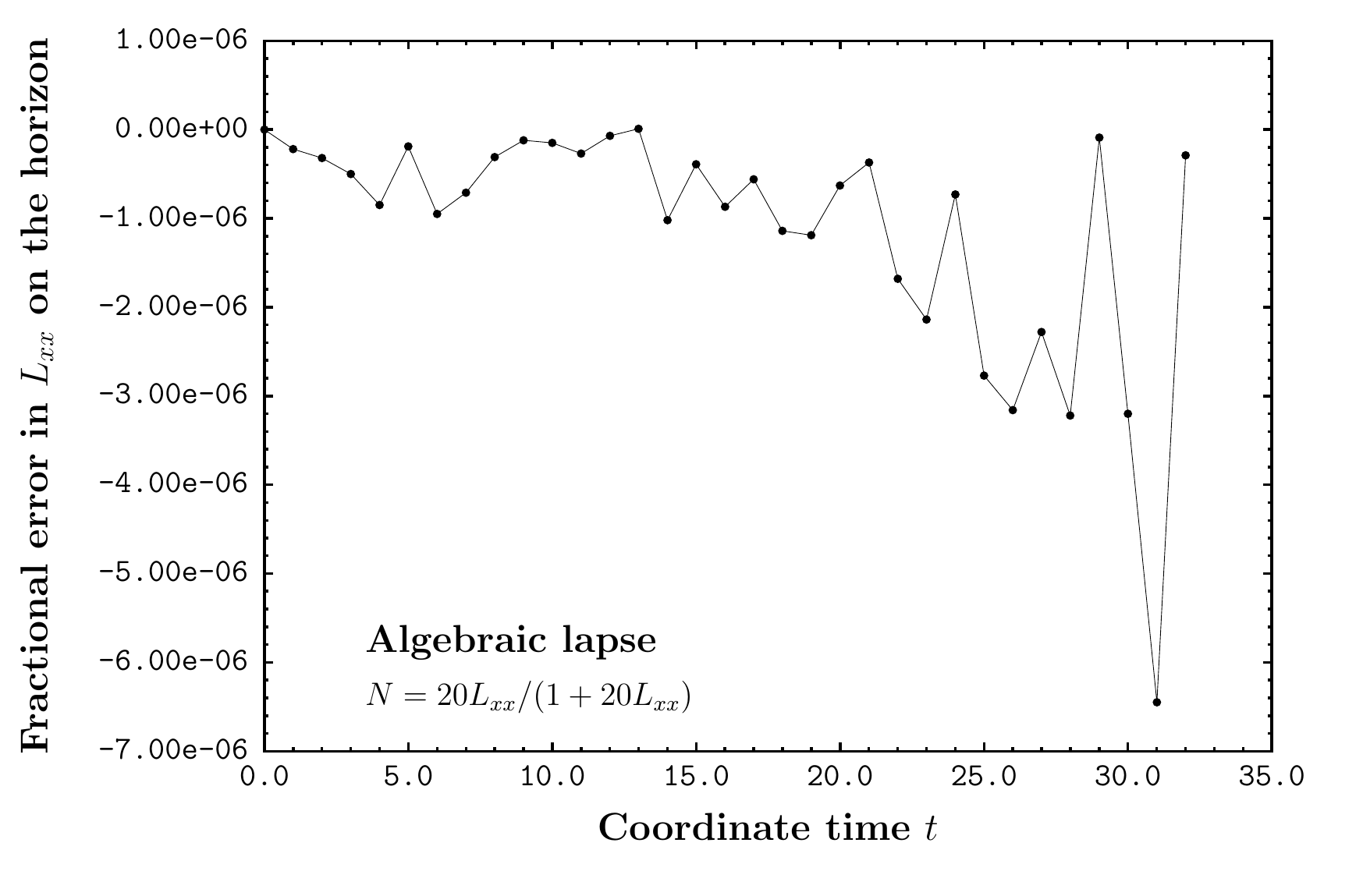}%
        {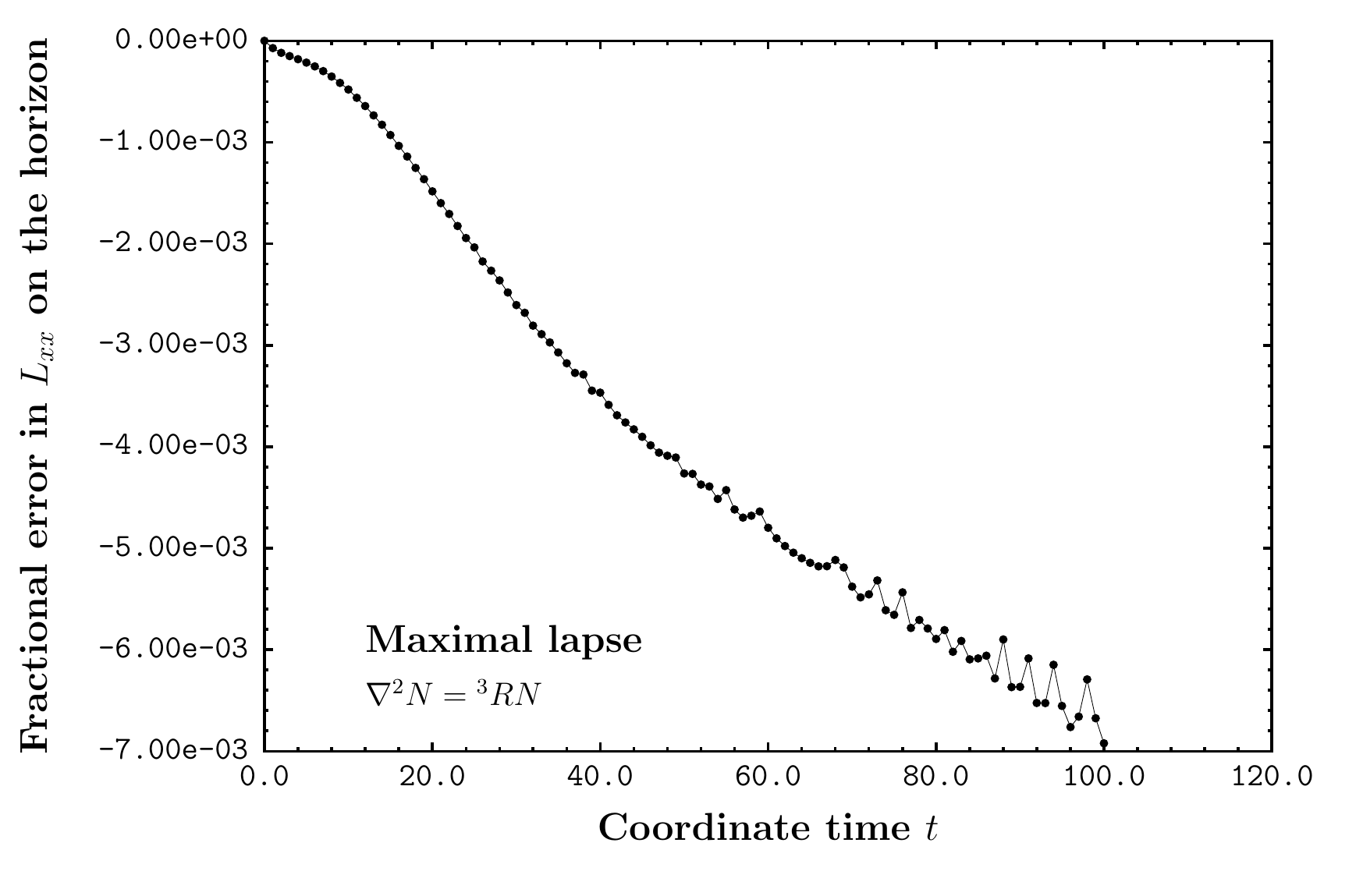}%
        {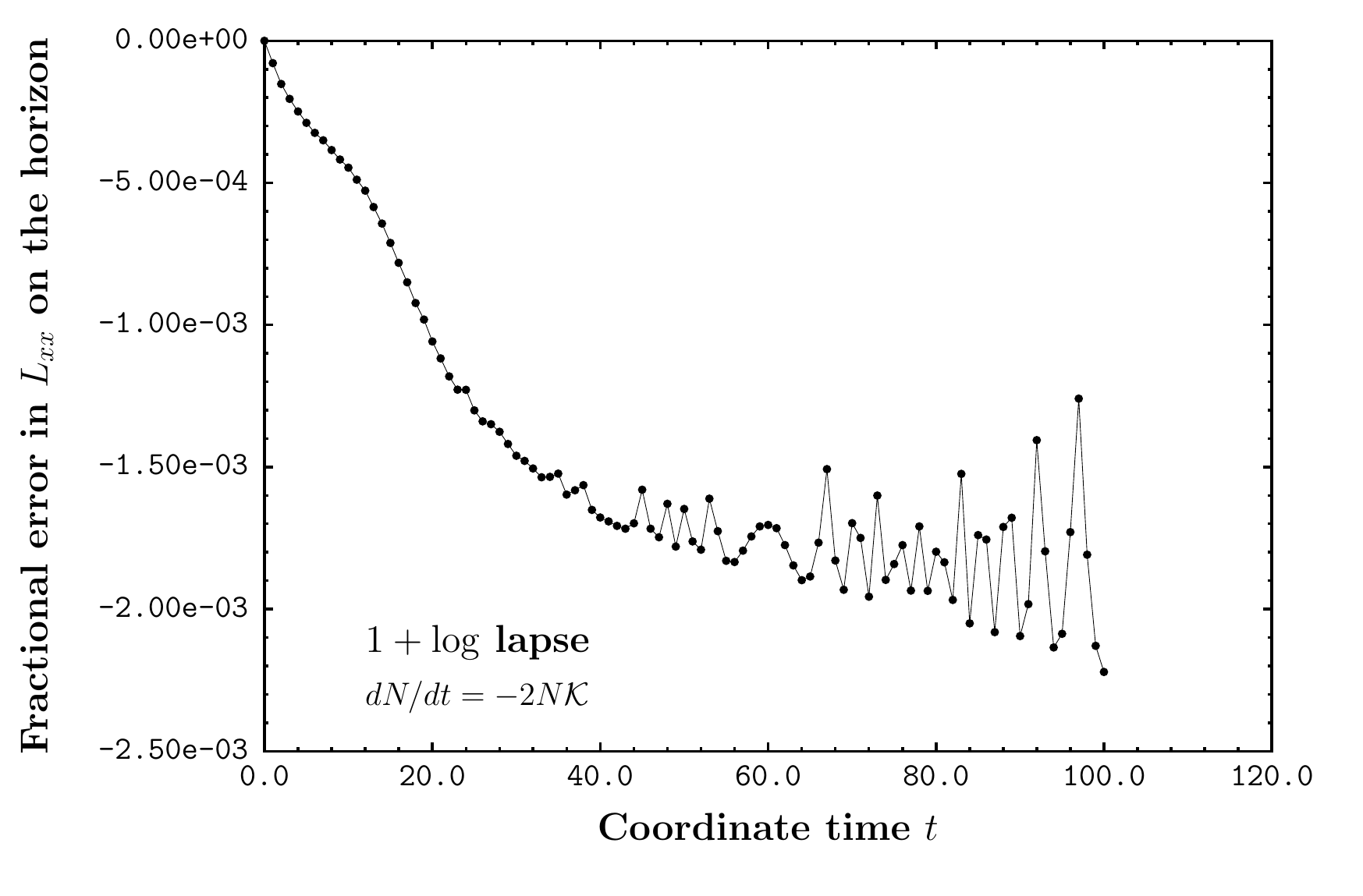}
\caption{\normalfont%
In the absence of gravitational radiation the area of the horizon should remain constant. It
follows that the $\Lxx$ should be constant on the horizon. Here we plot the fractional
variation of $\Lxx$ on the horizon. The irregular behaviour of the plots for later times is
due, in part, to the difficulty in accurately locating the horizon.}
\label{fig:LxxHorizon}
\end{figure}

\begin{figure}[t]
\FigQuad{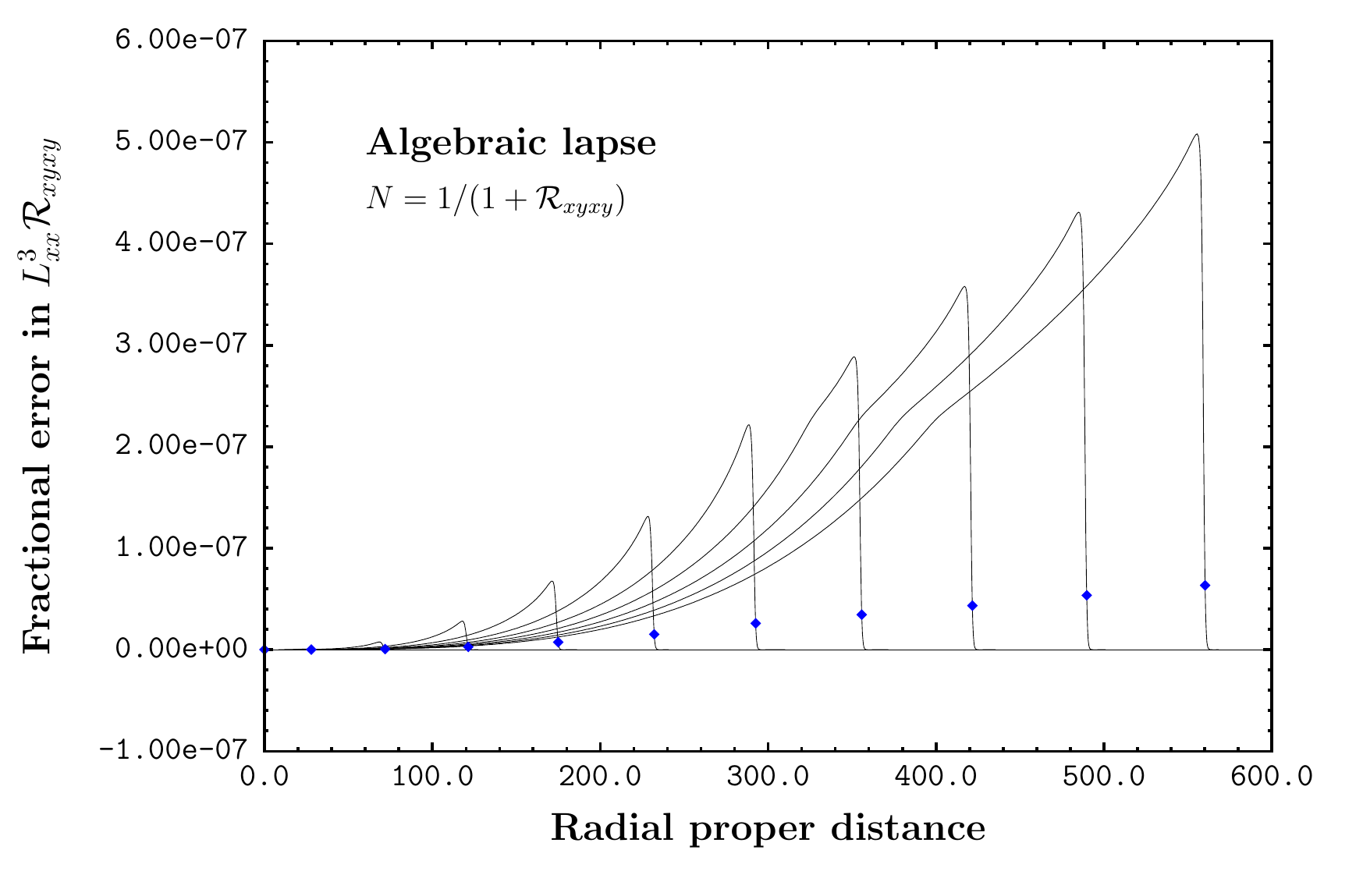}%
        {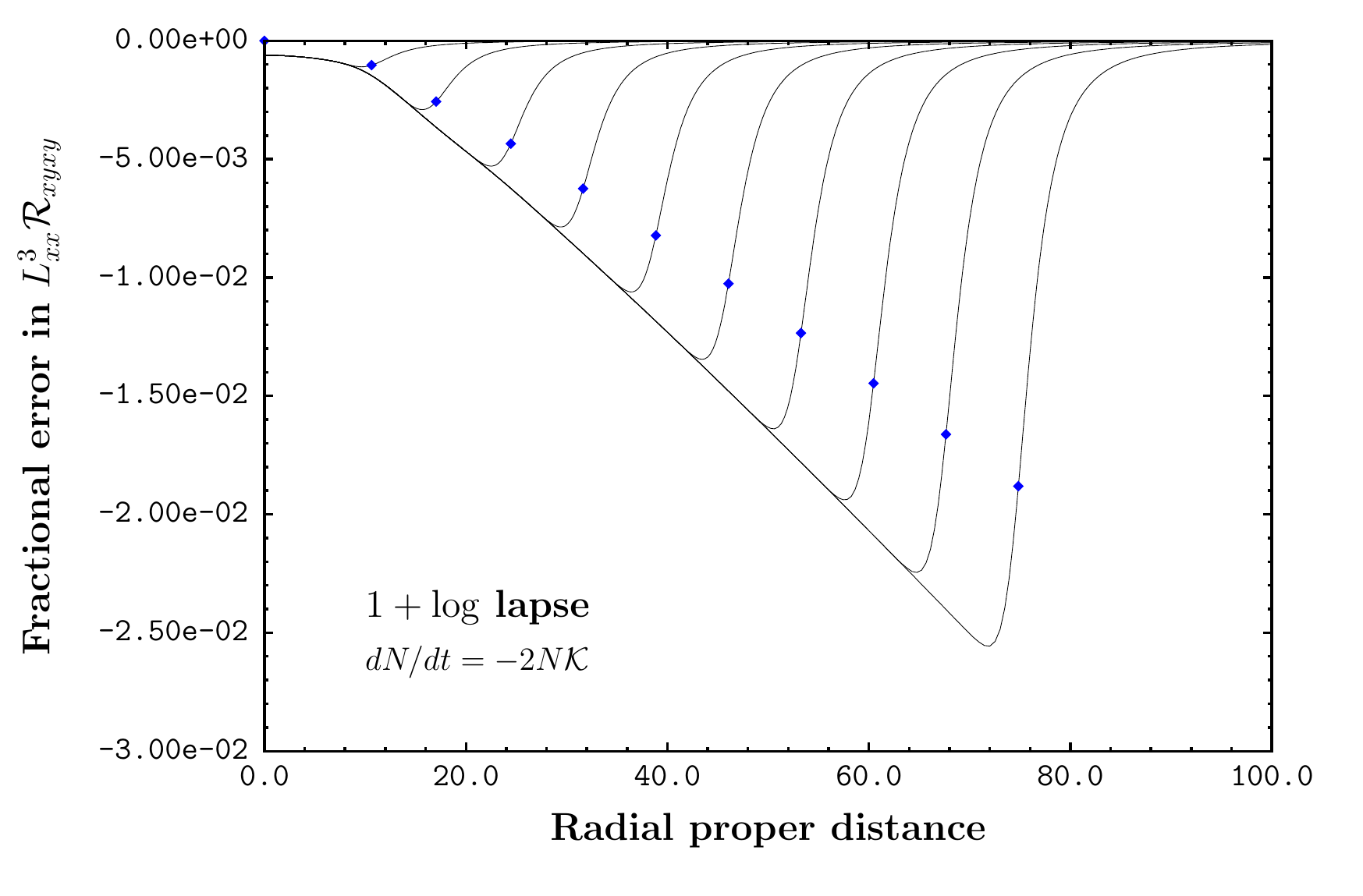}%
        {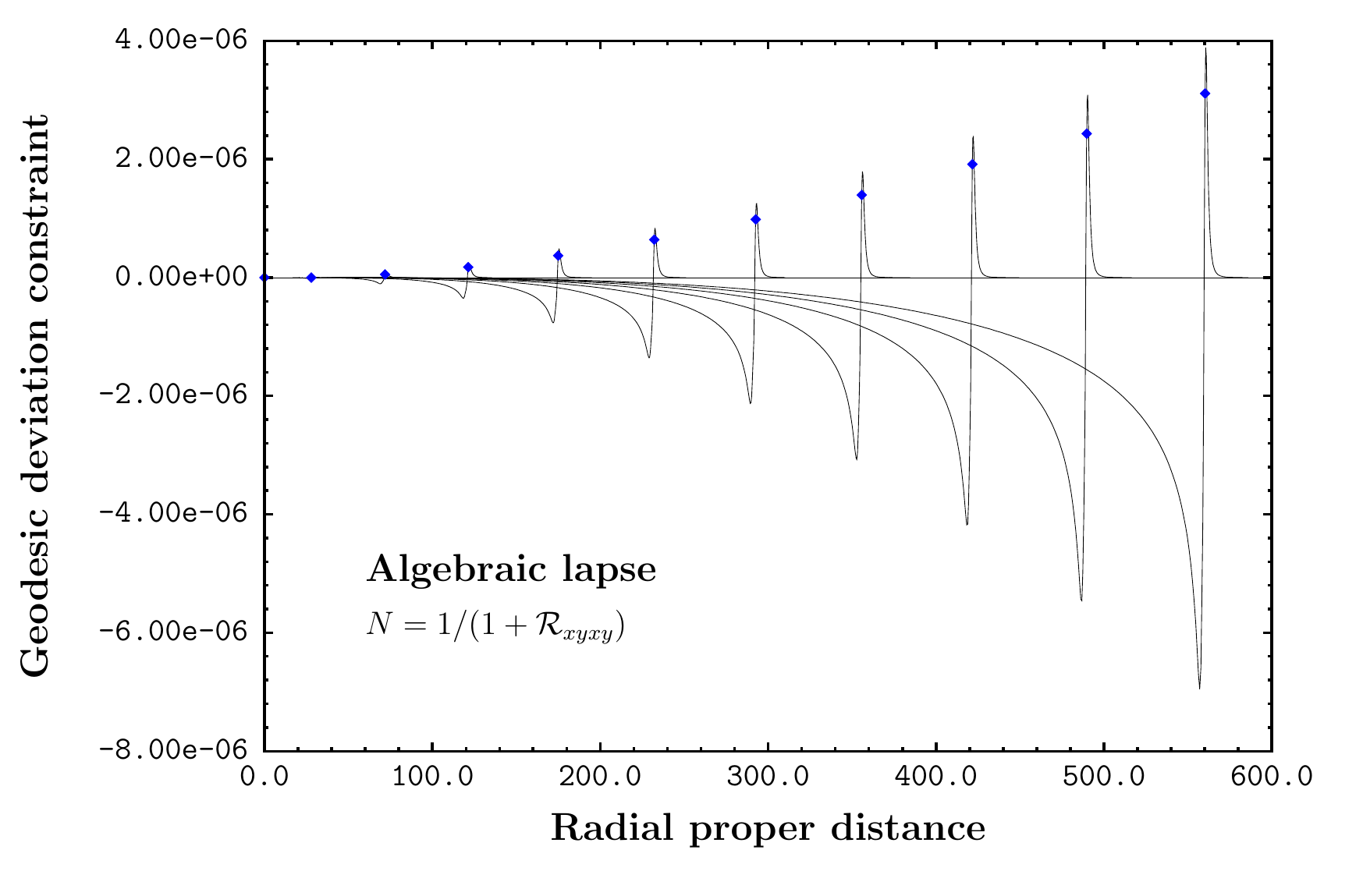}%
        {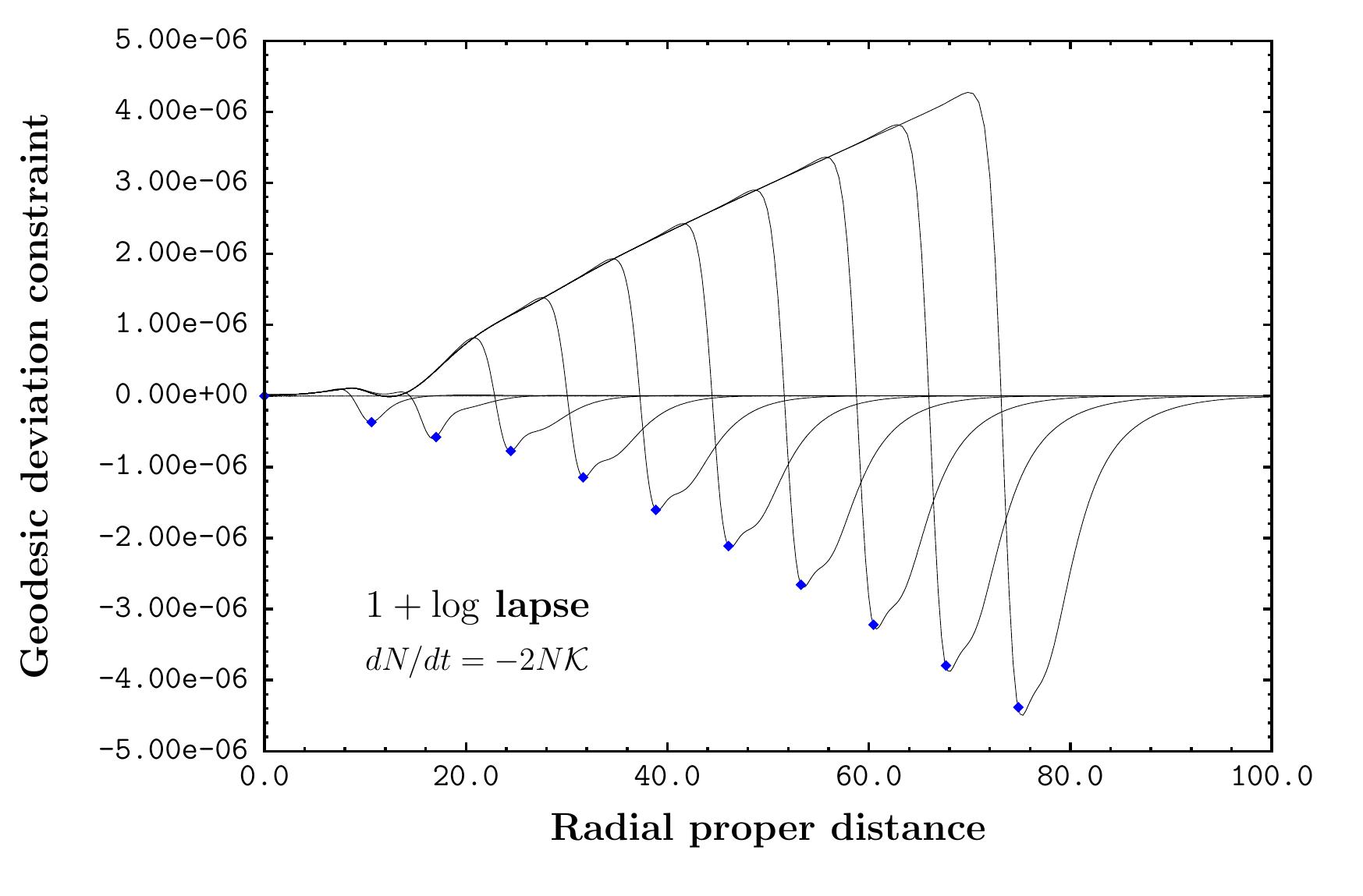}
\caption{\normalfont%
According to equation (\ref{eqn:ConserveLx3Rx}) the quantity $\Lxx^3\Rx$ should be conserved.
In the top row we display the relative error, defined by $1-C(t)/C(0)$ with
$C(t)=\Lxx^3(t)\Rx(t)$, for two choices of lapse. The errors for the $1+\log$ slicing are
much larger than those for the algebraic slicing which we attribute to the action of
the artificial viscosity terms. In the bottom row we display the momentum constraint for the
same pair of slicings. This shows a slow growth in the momentum constraint over time (judging
by the peaks in the plots, the growth appears to be linear in time).}
\label{fig:Lx3RxErr}
\end{figure}

\clearpage


\providecommand{\href}[2]{#2}\begingroup\raggedright\endgroup


%
%

%
%

%

\end{document}